\documentclass[pra, nofootinbib]{revtex4}
\usepackage{amssymb, color, amsmath, url, amstext,hyperref}
\usepackage[normalem]{ulem}
\usepackage{makeidx}
    \makeindex
\usepackage{times}
\usepackage{tabularx}
\usepackage{booktabs}
\usepackage{epsfig}

\newcommand{\Bra}[1]{\ensuremath{\langle#1|}}
\newcommand{\Ket}[1]{\ensuremath{|#1\rangle}}

\begin{document}

\title{Lectures on Quantum Information\\Chapter 1: The separability versus entanglement problem}
\author{Sreetama Das\(^{1,2}\), Titas Chanda\(^{1,2}\), Maciej Lewenstein\(^{3, 4}\), Anna Sanpera\(^{4, 5}\), Aditi Sen(De)\(^{1,2}\), and Ujjwal Sen\(^{1,2}\)}
\affiliation{$^1$Harish-Chandra Research Institute, Chhatnag Road, Jhunsi, Allahabad 211019, India \\
$^2$Homi Bhabha National Institute, Training School Complex, Anushaktinagar, Mumbai - 400094, India \\
\(^3\)ICFO-Institut de Ci\`encies Fot\`oniques, The Barcelona Institute of Science and Technology, 08034 Castelldefels (Barcelona), Spain\\
\(^4\) ICREA, Passeig de Lluis Companys 23, E-08010 Barcelona, Spain \\
\(^5\) Departament de F\'isica, Universitat Aut\`onoma de Barcelona, 08193 Bellaterra, Spain}

\maketitle



\tableofcontents





\section{Introduction}
Quantum theory, formalized in the first few decades of the 20\(^{\mbox{th}}\)century, contains elements that are radically different 
from the classical description of Nature. 
An important aspect in these fundamental differences
is the existence of quantum correlations in the quantum formalism. 
In the classical description of Nature, if a system is formed by  
different subsystems, 
complete knowledge of the whole system
implies that 
the sum of the information of the subsystems makes up the complete information for the whole
system. 
This is no longer true in the quantum formalism. In the quantum world, 
there exist states of composite systems 
for which we might have the complete knowledge, while our knowledge about
the subsystems might be completely random. In technical terms, one can have pure quantum states of a two-party system, whose local states are completely mixed.
One may reach some paradoxical conclusions if one applies a classical
description to states which have characteristic quantum signatures.

During the last two decades, it has been realized that these 
fundamentally nonclassical  states, also denoted as ``entangled states'',
can provide us with something else than paradoxes. 
They may be \emph{used} to perform tasks that cannot be achieved
with classical states.  
As benchmarks of this turning point in our view of such nonclassical states, 
one might mention  the spectacular discoveries of 
(entanglement-based) quantum cryptography (1991) \cite{Ekert},  quantum dense coding (1992) \cite{BW}, and 
quantum teleportation (1993)
\cite{BBCJPW}. 


In this chapter, we will consider both bipartite and multipartite composite systems. We will 
define formally what entangled states are, present 
some important criteria to discriminate entangled states from separable ones, and show
how they can be classified according to their capability to perform some precisely defined
tasks. 
Our knowledge in the subject of entanglement is still far from complete, 
although significant progress has been made in the recent years and 
very active research is currently underway. We will consider multipartite quantum states (states of more than two parties) in Section \ref{sec:multi}, until then, we consider only bipartite quantum states.

\section{Bipartite pure states: Schmidt decomposition}

Consider a bipartite system in a shared pure state. The two parties in possession of the system are traditionally denoted as Alice (A) and Bob (B),  who can be
located in distant regions. 
Let Alice's physical system be described 
by the Hilbert space \(\mathcal{H}_A\) and that of Bob by \(\mathcal{H}_B\). Then the joint
physical system of Alice and Bob is described by the tensor product Hilbert space 
\(\mathcal{H}_A \otimes \mathcal{H}_{B}\).

\noindent \textbf{Def. 1} \emph{Product and entangled pure states:}\index{Bipartite pure states}\\
\emph{
A pure state, i.e. a projector \(|\psi_{AB}\rangle \langle \psi_{AB}|\)
on a vector $|\psi_{AB}\rangle \in {\cal H}_A \otimes {\cal H}_B $, is a product state if the states of 
local subsystems are also pure states, that is, if  $|\psi_{AB}\rangle= |\psi_A\rangle \otimes |\psi_B\rangle$.
However, there are states that cannot be written in this form. 
These states are called entangled states.}

An example of an entangled state is the well-known singlet state \(\Ket{\Psi^-}=(\left|01\right\rangle - \left|10\right\rangle)/\sqrt{2}\), where 
\(|0\rangle\) and \(|1\rangle\) are two orthonormal states. 
Operationally, product states correspond  to those states, that can be locally prepared by Alice and Bob 
at two separate locations. 
Entangled states can, however, be prepared only after the particles of Alice and Bob have interacted either directly or
by means of an ancillary system. The second option is necessary due to the existence of the phenomenon of entanglement swapping \cite{ZZHE}.
A very useful representation, only valid for pure bipartite states, is the, so-called, Schmidt representation:\\
\textbf{Theorem 1} \emph{Schmidt decomposition:}\index{Schmidt decomposition}\\ 
\emph{Every $|\psi_{AB}\rangle \in {\cal H}_{A} \otimes {\cal H}_{B}$ can be represented in an
appropriately chosen basis as
\begin{equation}
|\psi_{AB}\rangle = \sum _{i=1}^{M} a_i |e_i\rangle \otimes |f_i\rangle,
\label{eq:schmidt}
\end{equation}
where $|e_i\rangle$ ($|f_i\rangle$) form a part of an orthonormal basis in
${\cal H}_{A}$ (${\cal H}_{B}$), $a_i > 0$, and  $\sum_{i=1}^{M} a_i^2=1$, where $M\le dim{\cal H}_{A}, dim{\cal H}_{B}$}.\\

The positive numbers \(a_i\) are known as the Schmidt coefficients and the vectors $\Ket{e_i} \otimes \Ket{f_i}$ as the Schmidt vectors of \(|\psi_{AB}\rangle\). Note that product pure states correspond to those
states, whose Schmidt decomposition has one and only one Schmidt coefficient. 
If the decomposition has more than one Schmidt coefficients, the state is entangled. 
Notice that the squares of the Schmidt coefficients of a pure bipartite state \(|\psi_{AB}\rangle\) are the
eigenvalues of both the reduced density matrices \(\rho_A\) (\(= \mbox{tr}_B (\Ket{\psi_{AB}} \Bra{\psi_{AB}}) \)) and 
 \(\rho_B\) (\(= \mbox{tr}_A (\Ket{\psi_{AB}} \Bra{\psi_{AB}})\)) of \(|\psi_{AB}\rangle\). The last fact gives us an easy method to find the Schmidt coefficients and the Schmidt vectors.



\section{Bipartite mixed states: Separable and entangled states}

As discussed in the last section, the question whether a given pure bipartite state is separable or entangled is 
straightforward.  One has just to check if the reduced density matrices 
are pure. This condition is equivalent  to the fact that a bipartite pure state has a 
single Schmidt coefficient. The determination of separability for mixed states is much harder, 
and currently lacks a complete answer, even in composite systems of dimension as low as \({\mathbb{C}}^2 \otimes {\mathbb{C}}^4\).

To reach a formal definition of separable and entangled states, consider the following preparation procedure 
of a bipartite quantum state between Alice and Bob. 
Suppose that Alice prepares her physical system in the state \(|e_{i}\rangle\) and Bob
prepares his physical system in the state \(|f_{i}\rangle\). 
Then, the combined state of their joint physical
system is given by:
\begin{equation}
|e_{i}\rangle \langle e_i |\otimes |f_{i}\rangle \langle f_i|.
\end{equation}
We now assume that they can communicate over a classical channel
(a phone line, for example). 
Then, whenever Alice prepares the state \(|e_{i}\rangle\) (\(i = 1,2, \ldots, K\)), which she does 
with probability \(p_i\), she communicates that to Bob, and correspondingly Bob prepares 
his system in the state \(|f_{i}\rangle\) (\(i = 1,2, \ldots, K\)). Of course, \(\sum_i p_i =1\) and $p_i \geq 0, \ \forall i$. The state that they 
prepare is then 
\begin{eqnarray}
\rho_{AB} = \sum_{i=1}^K  p_i |e_i\rangle \langle e_i| \otimes |f_i\rangle \langle f_i|.
\label{eqn:sepmixed}
\end{eqnarray}

\noindent \textbf{Def. 2} \emph{Separable and entangled mixed states:}\index{Bipartite mixed states}\\
\emph{A quantum state $\rho_{AB}$ is separable if and only if it can be represented as a convex combination of 
the product of projectors on local states as stated   in Eq. 
(\ref{eqn:sepmixed}). Otherwise, the state is said to be entangled.}

The important point to note here is that the state displayed in Eq. (\ref{eqn:sepmixed}) is the most general 
state that Alice and Bob will be able to prepare by local quantum operations and classical communication (LOCC) \cite{Werner}. 
In LOCC protocols, two parties Alice and Bob perform local quantum operations separately in their respective Hilbert spaces and they are allowed to communicate classical information about the results of their local operations. Let us make the definition somewhat more formal.\\
\noindent\textbf{Local operations and classical communication (LOCC):} \index{Local operations and classical communication (LOCC)} 
Suppose Alice and Bob share a quantum state $\rho_{AB}$ defined on the Hilbert space $\mathcal{H}_A \otimes \mathcal{H}_B$. Alice performs a quantum operation on her local Hilbert space $\mathcal{H}_A$, using a complete set of complete general quantum operations $\{A_i^{(1)}\}$, satisfying $\sum_i (A_i^{(1)})^{\dagger} A_i^{(1)} = I_A$, and sends her measurement result $i$ to Bob via a classical channel. Depending on the measurement result of Alice, Bob operates a complete set of general quantum operations $\{B_{ij}^{(1)}\}$, satisfying $\sum_j (B_{ij}^{(1)})^{\dagger} B_{ij}^{(1)} = I_B$ on his part belonging to the Hilbert space $\mathcal{H}_B$. This joint operation along with the classical communication is called one-way LOCC. Furthermore, Bob can send his result $j$ to Alice, and she can choose another set of local operations $\{A_{ijk}^{(2)}\}$, satisfying $\sum_k (A_{ijk}^{(2)})^{\dagger} A_{ijk}^{(2)} = I_A$, according to Bob's outcome. They can continue this process as long as required, and the entire operation is termed as LOCC, or two-way LOCC. The operators $I_A$ and $I_B$ are the identity operators on $\mathcal{H}_A$ and $\mathcal{H}_B$ respectively.


Entangled states cannot be prepared by two parties if only LOCC is allowed between them. To prepare such states, the physical systems
must be brought together to interact \footnote{\label{foot:ent_swap} Because of the phenomenon of 
entanglement swapping \cite{ZZHE}, one must suitably enlarge the notion of preparation of entangled 
states. So, an entangled state between two particles can be prepared if and only if, either the two particles 
(call them A and B) themselves came together to interact at a time in the past, or two \emph{other} particles (call them C 
and D) does the same, with C (D) having interacted beforehand with A (B).}.

The question whether a given bipartite state is separable or not turns out to be quite complicated. 
Among the difficulties, we notice that for an arbitrary state \(\rho_{AB}\), there is no stringent 
bound on the value of \(K\) in Eq. (\ref{eqn:sepmixed}), which is only limited by the Caratheodory theorem to  
be \(K \leq (\dim {\cal H} )^2\) with 
${\cal H} = {\cal H}_A \otimes {\cal H}_B$ (see \cite{Pawelbound, karnas00}).
Although the general answer to the separability problem still eludes us, there has been significant progress in recent years, and we will review some such directions in the following sections.

\section{Operational entanglement criteria}\index{Entanglement criteria}

In this section, we will introduce some operational entanglement criteria for bipartite quantum states.
In particular, we will discuss the partial transposition criterion \cite{PeresPPT, HorodeckiPPT}, the majorization criterion \cite{NielsenKempe}, the cross-norm or realignment criterion \cite{CCN1, CCN2, CCN3}, and the covariance matrix criterion \cite{cmc1, cmc2}. 
There exist several other criteria 
(see e.g. Refs. \cite{reduction, Doherty, other_criteria1, other_criteria2, other_criteria3, other_criteria4}), which  will not be discussed here. 
However note that, up to now, a necessary and sufficient criterion for detecting 
entanglement  of an arbitrary given mixed state is still lacking.

\subsection{Partial Transposition}\index{Partial transposition}
\label{sec:PPT}
\noindent \textbf{Def. 3} \emph{Let \(\rho_{AB}\) be a bipartite density matrix, and let us express it as 
\begin{equation}
\rho_{AB} = \sum_{\substack{1 \leq i,j \leq d_A \\ 1 \leq \mu, \nu \leq d_B}}  a_{ij}^{\mu\nu} (|i\rangle\langle j|)_A \otimes (|\mu\rangle\langle \nu|)_B,
\label{eq:mixed_state}
\end{equation}
where \(\{|i\rangle\}\) (\(\{|\mu\rangle\}\)) is a set of real orthonormal vectors in \({\cal H}_A\) (\({\cal H}_B\)), with $d_A = \dim {\cal H}_A$ and $d_B = \dim {\cal H}_B$. 
The partial transposition, \(\rho_{AB}^{T_A}\), of  \(\rho_{AB}\)
with respect to subsystem \(A\), 
is defined as}
\begin{equation}
\label{eq_partial_trans}
\rho_{AB}^{T_A} = 
\sum_{\substack{1 \leq i,j \leq d_A \\ 1 \leq \mu, \nu \leq d_B}} a_{ij}^{\mu\nu} (|j\rangle\langle i|)_A \otimes (|\mu\rangle\langle \nu|)_B.
\end{equation}

A similar definition exists for the partial transposition of \(\rho_{AB}\) with respect to Bob's subsystem. Notice that 
$\rho_{AB}^{T_{B}}\ =(\rho_{AB}^{T_{A}})^{T}$.
Although the partial transposition depends upon the choice of the basis in
which \(\rho_{AB}\) is written, its eigenvalues are basis independent.
We say that a state has  positive partial transposition (PPT) \index{Positive partial transposition},
whenever \(\rho_{AB}^{T_A} \geq 0\), i.e. the eigenvalues of \(\rho_{AB}^{T_{A}}\) are non-negative. 
Otherwise, the state is said to be non-positive under  partial transposition (NPT).

\noindent \textbf{Theorem 2} \cite{PeresPPT}\\
\emph{If  a state $\rho_{AB}$ is separable, then $\rho_{AB}^{T_{A}}\ \ge\ 0$ and
$\rho_{AB}^{T_{B}}\ =\left(\rho_{AB}^{T_{A}}\right)^{T}\ \ge\ 0$.}

\noindent Proof:\\
Since $\rho_{AB}$ is separable, it can be written as
\begin{eqnarray}
\rho_{AB} &
 = \sum_{i=1}^{K}\ p_i |e_i \rangle \langle e_i |\otimes |f_i\rangle \langle f_i | \ge 0.
\end{eqnarray}
Now performing the partial transposition w.r.t. A, we have 
\begin{eqnarray}
\rho_{AB}^{T_{A}} &=& \sum_{i=1}^{K}\ p_i\left(|e_i\rangle \langle e_i |
\right)^{T_{A}}\otimes |f_i \rangle \langle f_i |\nonumber\\
&=& \sum_{i=1}^{K}\ p_i |e_i^*\rangle \langle e_i^*| \otimes | f_i\rangle \langle f_i | \ge 0.
\end{eqnarray}
Note that in the second line, we have used the fact that $A^{\dagger}=\left(A^*\right)^{T}$. \(\square\)

The \emph{partial transposition criterion},
\index{Partial transposition criterion} for detecting entanglement is simple: Given a bipartite 
state \(\rho_{AB}\), find the eigenvalues of any of its partial transpositions. A negative eigenvalue immediately 
implies that the state is entangled. Examples of states for which the partial transposition has negative eigenvalues include
the singlet state. 

\begin{figure}
\includegraphics[scale=0.3]{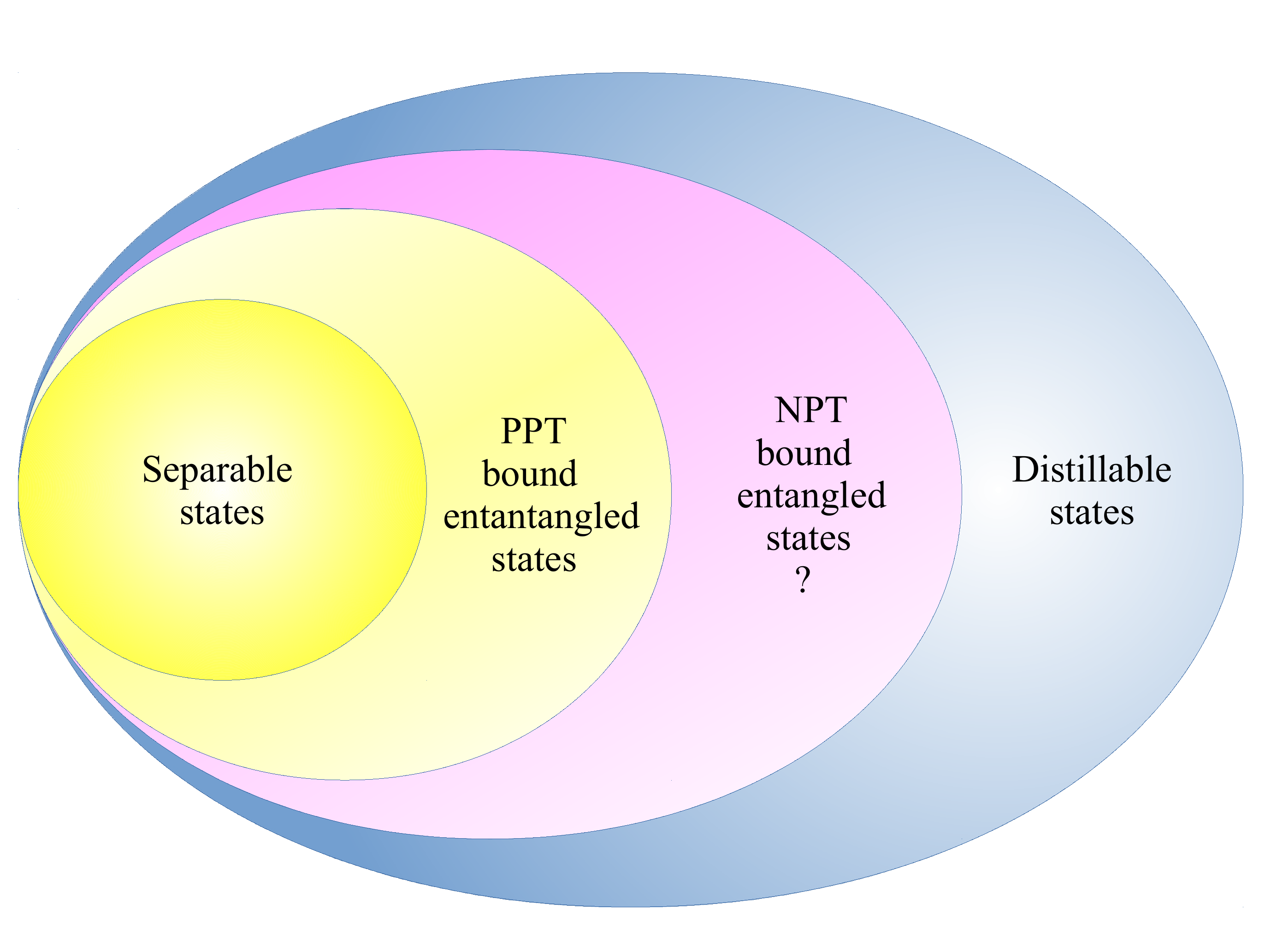}
\caption{The structure of the state space in light of the partial transposition criteria and distillability. Separable and as well as PPT states form convex sets (while NPT states do not). It also shows the conjectured NPT bound entangled states. It is not clear whether the set of non-distillable states is convex.}
\label{fig:phase}
\end{figure}

The partial transposition criterion allows to detect in a straightforward manner all entangled states that are
NPT states. This is a huge class of states. However, it turns out that there exist 
PPT states which are not separable, as pointed out in Ref. \cite{Pawelbound} (see also \cite{Horodeckibound}). 
Moreover, the set of PPT entangled states is not a set of measure zero
\cite{koto-volume-re}. 
It is, therefore, important to have further independent criteria 
of entanglement detection which permits to detect entangled PPT states. It is worth mentioning here that 
PPT states which are entangled, form the only known examples of the ``bound entangled states'' (see Refs. \cite{Horodeckibound,biyog} for details). Bound entangled states \index{Bound entangled states} of bipartite quantum states are the states that cannot be \emph{distilled} i.e., converted to singlet states under LOCC \cite{bennett_dist, bennett_eof2}, with other entangled states being distillable. We will talk about distillation of quantum states later in this chapter in a bit more detail. Although as yet not found, it is conjectured that there also exist NPT bound entangled states \cite{biyog}. Note also that both separable as well as PPT states form convex sets. Figure \ref{fig:phase} depicts the structure of the state space with respect to the partial transposition criteria and distillability.

Theorem 2 is a necessary condition of separability in any arbitrary dimension. However, 
for some special cases, the partial transposition criterion is both a necessary and a sufficient condition for
separability:\\
\textbf{Theorem 3} \cite{HorodeckiPPT}\\
\emph{In ${\mathbb{C}}^2 \otimes {\mathbb{C}}^2$ or ${\mathbb{C}}^2 \otimes {\mathbb{C}}^3$,
a state $\rho$ is separable if and only if $\rho^{T_{A}}\ \ge\ 0$}.

As mentioned above, PPT bound entangled states exist. However, as the Theorem 3 shows, they can exist only in dimensions higher than $\mathbb{C}^2 \otimes \mathbb{C}^2$ and $\mathbb{C}^2 \otimes \mathbb{C}^3$.










\subsection{Majorization}\index{Majorization}

The partial transposition criterion, although powerful, is not able to detect entanglement in 
a finite volume of states. It is, therefore, 
interesting to discuss other independent criteria. The majorization criterion\index{Majorization criterion}, to be 
discussed in this subsection, has been shown to be \emph{not} more powerful in detecting entanglement. 
We choose to discuss it here, mainly because it has independent roots. Moreover, it reveals a very interesting 
thermodynamical property of entanglement.

Before presenting the criterion,
we present a definition of majorization \cite{majorizationBhatia}.\\
\textbf{Def. 4} \emph{Let \(x = (x_1, x_2, \ldots, x_d )\), and \(y  = (y_1, y_2, \ldots, y_d )\) be two probability distributions, arranged in
decreasing order, i.e. \(x_1 \geq x_2 \geq \ldots \geq x_d\) and \(y_1 \geq y_2 \geq \ldots \geq y_d\). 
Then we define ``\(x\) majorized by \(y\)'', 
denoted as \(x\prec y\), as} 
\begin{equation}
 \sum_{i=1}^l x_i \leq \sum_{i=1}^l y_i,  
 \end{equation}
\emph{where \( l = 1,2, \ldots d-1\), and equality holds when \(l = d\)}.

\noindent \textbf{Theorem 4} \cite{NielsenKempe}\\
\emph{If a state \(\rho_{AB}\) is separable,  then}
\begin{equation}
\label{dibakar}
\lambda (\rho_{AB}) \prec \lambda (\rho_A ), \quad  and \quad \lambda (\rho_{AB}) \prec \lambda (\rho_B ), 
\end{equation}
\emph{where \(\lambda (\rho_{AB})\) is the set of eigenvalues of \( \rho_{AB} \), and 
\(\lambda ( \rho_A )\) and \( \lambda ( \rho_B ) \) are the sets of eigenvalues
of the reduced density matrices of the state \(\rho_{AB}\), and where all the sets are arranged in decreasing order.}

The majorization criterion: Given a bipartite state, it is entangled if Eq. (\ref{dibakar}) is violated. 
However, it was shown in Ref. \cite{Hiroshima}, that a state that is not detected by the 
positive partial transposition criterion, will not be detected by the majorization criterion either. 
Nevertheless, the criterion has other important implications. We will now discuss one such.



Let us reiterate an interesting fact about the singlet state: The global state is pure, while the local states are 
completely mixed. In particular, this implies that the von Neumann entropy \footnote{The von Neumann entropy of a state \(\rho\) is 
\(S(\rho) = -\mbox{tr}\rho \log_2 \rho\).}
 of the singlet is lower than 
the von Neumann entropies of either of the local states.
Since the von Neumann entropy can be used to quantify disorder in a given state, there exist global states whose disorder is lower than the any of the local states.
This is a nonclassical fact as for two classical random variables, the 
Shannon entropy \footnote{The Shannon entropy of a random variable \(X\), taking up values \(X_i\), with probabilities \(p_i\),
is given by \(H(X) = H(\{p_i\}) = - \sum_i p_i \log_2 p_i\).} of the joint distribution cannot be smaller than that of either. 
In Ref. \cite{horo_ent_sep}, it was shown that a similar fact is true for separable states:\\
\textbf{Theorem 5}\\
\emph{If a state \(\rho_{AB}\) is separable, 
\begin{equation}
\label{chandrima}
S (\rho_{AB}) \geq S (\rho_A ), \quad  and \quad S (\rho_{AB}) \geq S (\rho_B ). 
\end{equation}
}
Although the von Neumann entropy is an important notion for quantifying disorder, the theory of majorization 
is a more stringent quantifier \cite{majorizationBhatia}:
For two probability distributions \(x\) and \(y\), \(x \prec y\) if and only if \(x =  Dy\), where 
\(D\) is a doubly stochastic matrix \footnote{A matrix \(D= (D_{ij})\) is said to be doubly stochastic, if \(D_{ij}\geq 0\), and 
\(\sum_i D_{ij} = \sum_j D_{ij} =1\).}. Moreover, \(x \prec y\) implies that \(H(\{x_i\}) \geq H(\{y_i\})\).
Quantum mechanics therefore allows the existence of states for which global disorder is greater than local disorder even in the 
sense of majorization.

 A density matrix that satisfies Eq. (\ref{dibakar}), 
automatically satisfies Eq. (\ref{chandrima}). In this sense, Theorem 4 is a generalization of Theorem 5.

\subsection{Cross-norm or matrix realignment} \index{Cross-norm or matrix realignment criterion}
The cross-norm or matrix realignment criterion \cite{CCN1, CCN2, CCN3} provides another way to delineate separable and entangled states, and more importantly, can successfully detect various PPT entangled states. 
There are various ways to formulate this criterion.
Here we  present a formulation given in Ref. \cite{CCN3} as Corollary 18.

 A density matrix $\rho_{AB}$ on a Hilbert space $\mathcal{H}_{A} \otimes \mathcal{H}_{B}$, where $d_{A}$ and $d_{B}$ are the dimensions of the Hilbert spaces $\mathcal{H}_A$ and $\mathcal{H}_B$ respectively, can be written as 
\begin{equation}
 \rho_{AB} = \sum_{\substack{1 \leq i,j \leq d_A \\ 1 \leq \mu, \nu \leq d_B}} a_{ij}^{\mu \nu} (\Ket{i}\Bra{j})_A \otimes (\Ket{\mu}\Bra{\nu})_B 
 = \sum_{k,l} \xi_{kl} \tilde{G}_{k}^{A} \otimes \tilde{G}_{l}^{B}.
\label{eq:ccn1}
\end{equation}  
We have used the same notations as in Eq. (\ref{eq:mixed_state}), except that we have added zeros to the tensor $a_{ij}^{\mu \nu}$, so that the indices run until the dimensions of the Hilbert spaces.
 $\{\tilde{G}_{k}^{A}\} = \{\Ket{i}\Bra{j}\}$ and $\{\tilde{G}_{l}^{B}\} = \{\Ket{\mu}\Bra{\nu}\} $ are complete sets of orthonormal Hermitian operators on the Hilbert spaces $\mathcal{H}_A$ and $\mathcal{H}_B$ respectively, with $ 1 \leq k \leq d_{A}^{2} $ and $ 1 \leq l \leq d_{B}^{2}$.
 Without loss of generality, we assume that $d_{A} \geq d_{B}$. After singular value decomposition of the matrix $\xi$, we have
\begin{equation}
\xi = U \Sigma V^{\dagger},
\end{equation} 
  where $U$ and $V$ are $ d_{A}^{2} \times d_{A}^{2} $ and $ d_{B}^{2} \times d_{B}^{2} $ dimensional unitary matrices respectively, and $ \Sigma $ is a $ d_{A}^{2} \times d_{B}^{2} $ dimensional diagonal matrix. Denoting the $k$th column vector of $ U $ and $ V $ by $ \Ket{u_k} $ and $ \Ket{v_k} $, the above expression becomes
  \begin{equation}
  \xi = \sum_{k = 1}^{d_B^2} \lambda_{k} \Ket{u_{k}} \Bra{v_{k}},
  \end{equation}
  where $ \lambda_{k}$ are the diagonal elements of $\Sigma$. So, we have the matrix elements of $\xi$ as
  \begin{equation}
  \xi_{kl} = \lambda_{k} \delta_{kl}.
  \label{eq:ccn2}
  \end{equation}
If $G_k^A$  and $G_l^B$ are the matrix representations of $\tilde{G}_k^A$ and $\tilde{G}_l^B$ in $\{\Ket{u_k}\}$ and $\{\Ket{v_k}\}$ basis respectively, then using Eqs. (\ref{eq:ccn1}) and (\ref{eq:ccn2}), we obtain
  \begin{equation}
  \rho_{AB} = \sum_{k = 1}^{d_B^2} \lambda_{k} G_{k}^{A} \otimes G_{k}^{B}.
  \label{eq:ccn3}
  \end{equation}
Eq. (\ref{eq:ccn3}) can be interpreted as the Schmidt decomposition of the density matrix $\rho_{AB}$ in operator space, where the singular values  $\lambda_k$ are real and non-negative. The cross-norm or realignment criterion of separability is given by the following theorem:   

\noindent\textbf{Theorem 6} \cite{CCN3}\\
\emph {If a shared quantum state $\rho_{AB}$ is separable, then
  \begin{equation}
  \sum_{k} \lambda_{k} \leq 1,
  \end{equation}
 where $\lambda_{k}$ are the singular values of $\rho_{AB}$ as given in Eq. (\ref{eq:ccn3}). If the inequality is violated, one can conclude that $\rho_{AB}$ must be an entangled state.}

\subsection{Covariance matrix} \index{Covariance matrix}
There exist several other operational criteria in the literature to detect whether a quantum state is separable or entangled \cite{reduction, Doherty, other_criteria1, other_criteria2, other_criteria3, other_criteria4}. We will conclude this section by briefly illustrating one such separability criteria, known as the covariance matrix criterion \cite{cmc1, cmc2}, which provides a general framework to link and understand several existing criteria including the cross-norm or realignment criterion. Like the cross-norm or realignment criterion, this method can identify entangled state for which  the partial transposition criterion fails.
Before delving into the theory of the covariance matrix criterion, let us first discus the definition and properties of the covariant matrices.

\subsubsection{Definition and properties}
\noindent \textbf{Def. 5} \emph{Given a quantum state $\rho$ and a complete set of orthonormal observables $\{M_k\}$ on a $d$-dimensional Hilbert space, the $d^2 \times d^2$ covariant matrix $\gamma = \gamma(\rho,\{M_k\})$ and symmetrized covariant matrix $\gamma^S = \gamma^S(\rho,\{M_k\})$  are defined as
\begin{eqnarray}
\gamma_{ij} &=& \langle M_i M_j \rangle - \langle M_i \rangle \langle M_j \rangle, \\
\gamma^S_{ij} &=& \frac{\langle M_i M_j \rangle + \langle M_j M_i \rangle}{2} - \langle M_i \rangle \langle M_j \rangle,
\end{eqnarray}
where $\langle M \rangle = \mbox{tr}(\rho M)$ defines the expectation of the operator $M$ with respect to the state $\rho$.} 

The complete set of orthonormal observables $\{M_k\}$ has to satisfy the Hilbert-Schmidt orthonormality condition $\mbox{tr}(M_i M_j) = \delta_{ij},$ $\forall i,j = 1,2,...,d$.  One example for such a set of observables for the case
of single qubit in terms of the Pauli matrices, can be given by
\begin{equation}
M_1 = \frac{I}{\sqrt{2}}, \ M_2 = \frac{\sigma_x}{\sqrt{2}}, \ M_3 = \frac{\sigma_y}{\sqrt{2}},
\ M_4 = \frac{\sigma_z}{\sqrt{2}}.
\end{equation} 
In general, for the $d$-dimensional case, one can consider following matrices to form the complete set of orthonormal observables:
\begin{eqnarray}
X_i &=& \Ket{i}\Bra{i}, \  \ i = 1,2,...,d, \\
Y_{ij} &=&  \frac{1}{\sqrt{2}}(\Ket{i}\Bra{j} + \Ket{j}\Bra{i}), \ \ 1 \leq i < j \leq d, \\
Z_{kl} &=&  \frac{i}{\sqrt{2}}(\Ket{k}\Bra{l} - \Ket{l}\Bra{k}), \ \ 1 \leq k < l \leq d.
\end{eqnarray}

Let us now focus on the situation in which the Hilbert space is a tensor product $\mathcal{H}=\mathcal{H}_A \otimes \mathcal{H}_B$ of Hilbert spaces of two subsystems $A$ and $B$ with dimensions $d_A$ and $d_B$ respectively. We can consider the complete set of orthonormal observables in $A$ as $\{ A_k: k = 1,2...,d_A^2\}$ and  in $B$ as $\{ B_k: k = 1,2...,d_B^2 \}$, and construct a set of $d^2_A + d^2_B$ observables as $\{ M_k \} =\{ A_k \otimes I, I \otimes B_k\}$. Although this set is not complete, it can be utilized to define a very useful form of covariant matrices, known as the block covariant matrices.
The block covariant matrix for a given bipartite state $\rho_{AB}$ and orthonormal observables $\{M_k\}$ is defined as follows. 

\noindent\textbf{Def. 6} \emph{Let $\rho_{AB}$ be a quantum state of a bipartite system, and let $\{ M_k \} =\{ A_k \otimes I, I \otimes B_k\}$ be a set of orthonormal observables as stated above. Then, the block covariance
matrix $\gamma = \gamma(\rho_{AB},\{M_k\})$ is given in terms of its matrix elements as $\gamma_{i,j} = \langle M_i M_j \rangle - \langle M_i \rangle \langle M_j \rangle$, and has the  block structure
\begin{equation}
\gamma = 
\begin{pmatrix}
\mathcal{A} && \mathcal{X} \\
\mathcal{X}^{T} && \mathcal{B}
\end{pmatrix},
\label{eq:bcmc}
\end{equation} 
where  $\mathcal{A} = \gamma(\rho_A,\{A_k\})$ and  $\mathcal{B} = \gamma(\rho_B,\{B_k\})$ are the covariant matrices of the reduced subsystems, and $\mathcal{X}_{i,j} = \langle A_i \otimes B_j \rangle - \langle A_i \rangle \langle B_j \rangle$.} 

Similarly, we can define the symmetric version of the block covariance matrix by replacing $\mathcal{A}$ and $\mathcal{B}$ with their symmetrized counterparts, while keeping $\mathcal{X}$ unchanged.  Clearly, if $\rho_{AB} = \rho_A \otimes \rho_B$ is a product state, then its block covariant matrix reduces to the block diagonal form, $\gamma(\rho_{AB}, \{M_k\}) = \mathcal{A} \oplus \mathcal{B}$, as $\mathcal{X}_{i,j}$ become zero $\forall i,j$.

If $\rho$ is a pure state on a $d$-dimensional Hilbert space, then the corresponding covariance matrix $\gamma$ satisfies following properties:
\begin{enumerate}
\item The rank of $\gamma$ is equal to $d-1$.
\item The nonzero eigenvalues of $\gamma$ are equal to 1, hence $\mbox{tr}(\gamma)=d-1$.
\item $\gamma^2 = \gamma$.
\end{enumerate}
The corresponding symmetric covariance matrix $\gamma^S$ satisfies the following:
\begin{enumerate}
\item The rank of $\gamma^S$ is equal to $2(d-1)$.
\item The nonzero eigenvalues of $\gamma^S$ are equal to 1/2, hence $\mbox{tr}(\gamma^S)=d-1$.
\end{enumerate}
For mixed state $\rho$ on a $d$-dimensional Hilbert space, we have $\mbox{tr}[\gamma(\rho)] = d - \mbox(\rho^2)$, and the same for $\gamma^S(\rho)$.
The covariance matrix (symmetric and non-symmetric) also satisfies the concavity property, i.e., if $\rho = \sum_k p_k \rho_k$ is a convex combination of states $\rho_k$, then
\begin{equation}
\gamma(\rho) \geq \sum_k p_k \gamma(\rho_k).
\end{equation}

\subsubsection{Covariance matrix criterion for separability}\index{Covariance matrix criterion}

\noindent\textbf{Theorem 7}\cite{cmc1, cmc2} \\
\emph{Let $\rho_{AB}$ be a separable state on $\mathcal{H}_A \otimes \mathcal{H}_B$ and $A_k$ and $B_k$ be the orthonormal observables on the Hilbert spaces $\mathcal{H}_A$ and $\mathcal{H}_B$, with the latter having dimensions $d_A$ and $d_B$ respectively. Define $\{ M_k \} =\{ A_k \otimes I, I \otimes B_k\}$ as discussed previously. Then there exist pure states $\Ket{\psi_k}\Bra{\psi_k}$ for $A$, $\Ket{\phi_k}\Bra{\phi_k}$ for $B$, and convex weights $p_k$ such that if we define $\kappa_A = \sum_k p_k \gamma(\Ket{\psi_k}\Bra{\psi_k}, \{ A_{k'} \})$ and $\kappa_B = \sum_k p_k \gamma(\Ket{\phi_k}\Bra{\phi_k}, \{ B_{k'} \})$, the inequality
\begin{equation}
\gamma(\rho_{AB},\{M_k\}) \geq \kappa_A \oplus \kappa_B \Leftrightarrow
\begin{pmatrix}
\mathcal{A} && \mathcal{X} \\
\mathcal{X}^T && \mathcal{B}
\end{pmatrix}
\geq 
\begin{pmatrix}
\kappa_A && 0 \\
0 && \kappa_B
\end{pmatrix}
\end{equation}
holds. 
} \\
Proof: \\
Let us assume $\rho_{AB}$ be a separable state with the following pure state decomposition,
\begin{equation}
\rho_{AB} = \sum_k p_k (\Ket{\psi_k}\Bra{\psi_k} \otimes \Ket{\phi_k}\Bra{\phi_k}).
\end{equation} 
Then using the properties of covariance matrices, we get
\begin{eqnarray}
\gamma(\rho_{AB},\{M_k\})  &=& \gamma(\sum_k p_k (\Ket{\psi_k}\Bra{\psi_k} \otimes \Ket{\phi_k}\Bra{\phi_k}), \{M_{k'}\}) \nonumber \\
&\geq&  \sum_k p_k \gamma(\Ket{\psi_k}\Bra{\psi_k} \otimes \Ket{\phi_k}\Bra{\phi_k}, \{M_{k'}\}) \nonumber \\
&=& \sum_k p_k \{\gamma(\Ket{\psi_k}\Bra{\psi_k}, \{A_{k'}\}) \oplus \gamma(\Ket{\phi_k}\Bra{\phi_k}, \{B_{k''}\})\} \nonumber \\
&=& \kappa_A \oplus \kappa_B,
\end{eqnarray}
where we have used concavity property of the covariance matrices in the second line, and in the third line, we have used the fact that the block covariant matrix of a product state takes the block diagonal form. This theorem can also be proven for the symmetric covariant matrices in the same manner. Clearly, if there exist no such $\kappa_A$ and $\kappa_B$, the state $\rho_{AB}$ must be entangled.\(\square\)

Clearly, it is not evident from Theorem 7 that the covariance matrix criterion leads to an efficient and physically plausible operational indication for separability. The main problem is to identify possible $\kappa_A$ and $\kappa_B$, as this requires an optimization over all pure state decompositions of $\rho_{AB}$. Therefore, we now focus on the cases, where the above criterion can be used efficiently to identify entangled states, by stating several corollaries of the above theorem.

\noindent \textbf{Corollary 1} \\
\emph{Let
$
\gamma = 
\begin{pmatrix}
\mathcal{A} && \mathcal{X} \\
\mathcal{X}^{T} && \mathcal{B}
\end{pmatrix}
$
be the block covariance matrix of a bipartite state $\rho_{AB}$. Then, if $\rho_{AB}$ is separable, we have}
\begin{equation}
||\mathcal{X}||^2 \leq [1 - \mbox{tr}(\rho_A^2)][1 - \mbox{tr}(\rho_B^2)],
\end{equation}
\emph{where} $||A|| = \mbox{tr} \ \sqrt{A^{\dagger}A}$ \emph{is the matrix trace norm.}   

\noindent \textbf{Corollary 2} \\
\emph{Let $\rho_{AB}$ be a quantum state shared between two subsystems $A$ and $B$ with dimensions $d_A$ and $d_B$ respectively (with $d_A \leq d_B$),
$
\gamma^S = 
\begin{pmatrix}
\mathcal{A} && \mathcal{X} \\
\mathcal{X}^{T} && \mathcal{B}
\end{pmatrix}
$
be the corresponding symmetric covariance matrix, and let $\mathcal{J} = \{j_1,...,j_{d_A^2}\} \left(\subset \{1,...,d_B^2\}\right)$ be a set of $d_A^2$ distinct indices. Then if $\rho_{AB}$ is separable, we have}
\begin{equation}
2 \sum_{i=1}^{d_A^2} \sum_{j \in \mathcal{J}} |\mathcal{X}_{i,j}| \leq [1 - \mbox{tr}(\rho_A^2)] + [1 - \mbox{tr}(\rho_B^2)].
\end{equation}

Now we will give another operational entanglement criterion based on the Schmidt decomposition on operator space, then try to relate covariance matrix criterion with the cross-norm or realignment criterion. A general bipartite quantum state 
$\rho_{AB}$ on a Hilbert space $\mathcal{H}_{A} \otimes \mathcal{H}_{B}$, where $d_{A}$ and $d_{B}$ are the dimensions of the Hilbert spaces $\mathcal{H}_A$ and $\mathcal{H}_B$ respectively, can be written as
\begin{equation}
\rho_{AB} = \sum_{k=1}^{d_A^2}\sum_{l=1}^{d_B^2} \xi_{kl} \tilde{G}^A_k \otimes \tilde{G}^B_l,
\label{eq:sd_svd}
\end{equation}  
where $\xi_{kl}$ are real quantities, and $\{ \tilde{G}^A_i \}$ and $\{ \tilde{G}^B_j \}$ are complete sets of orthonormal Hermitian operators on the Hilbert spaces $\mathcal{H}_A$ and $\mathcal{H}_B$ respectively. As we have seen earlier, $\rho_{AB}$ in Eq. (\ref{eq:sd_svd}), can be written in the Schmidt decomposed-like form (in operator space) via the singular value decomposition as
\begin{equation}
\rho_{AB} = \sum_{k=1}^{d_B^2} \lambda_k G^A_k \otimes G^B_k,
\end{equation}
where singular values $\lambda_k$ are real and non-negative, and we have assumed that $d_A \geq d_B$.

\noindent \textbf{Corollary 3} \\
\emph{Let $\rho_{AB}$ be a separable quantum state shared between two subsystems $A$ and $B$. Then
\begin{equation}
2\sum_i |\lambda_i - \lambda_i^2 g_i^A g_i^B| \leq 2 - \sum_i \lambda_i^2[(g_i^A)^2 + (g_i^B)^2],
\label{eq:col3}
\end{equation}
where $g_i^{A(B)} = \mbox{tr} [ G_i^{A(B)}]$. 
}\\
Proof: \\
Let
$
\gamma^S = 
\begin{pmatrix}
\mathcal{A} && \mathcal{X} \\
\mathcal{X}^{T} && \mathcal{B}
\end{pmatrix}
$
be the symmetric block covariance matrix of the bipartite state $\rho_{AB}$.
Using the orthogonality of the matrices $\{G_i^A\}$ and $\{G_i^B\}$, one can deduce $\mathcal{X}_{i,j} = \left(\lambda_i - \lambda_i^2 g_i^A g_i^B\right)$. Further, one can get $\mbox{tr}(\rho_A^2) = \sum_i \lambda_i^2 (g_i^B)^2$, and $\mbox{tr}(\rho_B^2) = \sum_i \lambda_i^2 (g_i^A)^2$. Together with Corollary 2, we can prove the claim. $\square$

Now using the relations $|a-b| \geq |a| - |b|$ and $a^2 + b^2 \geq 2|ab|$ we can have,
\begin{eqnarray}
2\sum_i |\lambda_i - \lambda_i^2 g_i^A g_i^B| &\geq & 2\sum_i (\lambda_i - \lambda_i^2 |g_i^A g_i^B|), \\
2(1 - \sum_i \lambda_i^2 |g_i^A g_i^B|) &\geq & 2 - \sum_i \lambda_i^2[(g_i^A)^2 + (g_i^B)^2].
\label{eq:col32}
\end{eqnarray}
Using inequalities (\ref{eq:col3})-(\ref{eq:col32}), we get
\begin{eqnarray}
2\sum_i |\lambda_i - \lambda_i^2 g_i^A g_i^B| &\leq & 2 - \sum_i \lambda_i^2[(g_i^A)^2 + (g_i^B)^2] \nonumber \\
&\leq & 2(1 - \sum_i \lambda_i^2 |g_i^A g_i^B|) \nonumber \\
&=& (2 - 2\sum_i \lambda_i) + 2\sum_i (\lambda_i - \lambda_i^2 |g_i^A g_i^B|) \nonumber \\
&\leq & (2 - 2\sum_i \lambda_i) + 2\sum_i |\lambda_i - \lambda_i^2 g_i^A g_i^B| \nonumber \\
\Leftrightarrow \sum_i \lambda_i &\leq & 1.
\end{eqnarray}
This is the cross-norm or realignment criterion of separability mentioned earlier, which we get as a corollary of the covariance matrix criterion.

There are several other corollaries of the covariance matrix criterion, which enable one to efficiently detect entangled states in several cases. Moreover, the covariance matrix criterion can be improved by using local filtering operation \cite{local_filter}.  See Ref. \cite{cmc2} for details.

\section{Non-operational entanglement criteria}

In this section, we will discuss three further entanglement criteria. We will show how the Hahn-Banach theorem can be used 
to obtain ``entanglement witnesses''. We will also introduce the notion of positive maps, and present the entanglement 
criterion based on it. And finally we will present the range criterion of separability.
All three criteria are ``non-operational'', in the sense that they are not state-independent. 
Nevertheless, they provide important insight into the structure of the set of entangled states. Moreover, the concept of entanglement witnesses
can be used to detect entanglement experimentally, by performing only a few \emph{local} measurements, assuming some prior 
knowledge of the density matrix \cite{sakkhi12, sakkhiprl}.


\subsubsection{Technical Preface}
The following lemma and observation will be useful for later purposes.\\
\textbf{Lemma 1}\\
$\mbox{tr}(\rho_{AB}^{T_{A}}\sigma_{AB}) = \mbox{tr}(\rho_{AB}\sigma_{AB}^{T_{A}})$.

\vspace{0.5cm}
\noindent \textbf{Observation:}\\
The space of linear operators acting on ${\cal H}$ (denoted by ${\cal B}({\cal H})$)
is itself a Hilbert space\index{Hilbert space}, 
with the (Euclidean) scalar product
\begin{equation}
\langle A|B\rangle = \mbox{tr}(A^\dagger B) \qquad A,B \in {\cal B}({\cal H}).
\end{equation}
This scalar product is equivalent to writing $A$ and $B$ row-wise as vectors,
and scalar-multiplying them:
\begin{equation}
\mbox{tr}(A^\dagger B)  = \sum_{ij}A^\ast_{ij}B_{ij} = \sum_{k=1}^{(\dim {\cal H})^2} a_{k}^\ast b_k.
\end{equation}

\subsection{Entanglement Witnesses}

\subsubsection{Entanglement Witness from the Hahn-Banach theorem}

Central to the concept of entanglement witnesses, is the Hahn-Banach
theorem, \index{Hahn-Banach theorem} which we will present here limited to our situation and without
proof (see e.g. \cite{alt:1985} for a proof of the more general theorem):\\
\textbf{Theorem 8}\\
\emph{
Let $S$ be a convex compact set in a finite dimensional Banach space. 
Let $\rho$ be a point in the space with $\rho \not\in S$.
Then there exists a 
hyperplane}\footnote{A hyperplane is a linear subspace with 
dimension one less than the dimension of the space itself.}\index{Hyperplane} \emph{that separates $\rho$ from $S$.}

\begin{figure}[htbp]
\begin{center}
\includegraphics[scale=0.3]{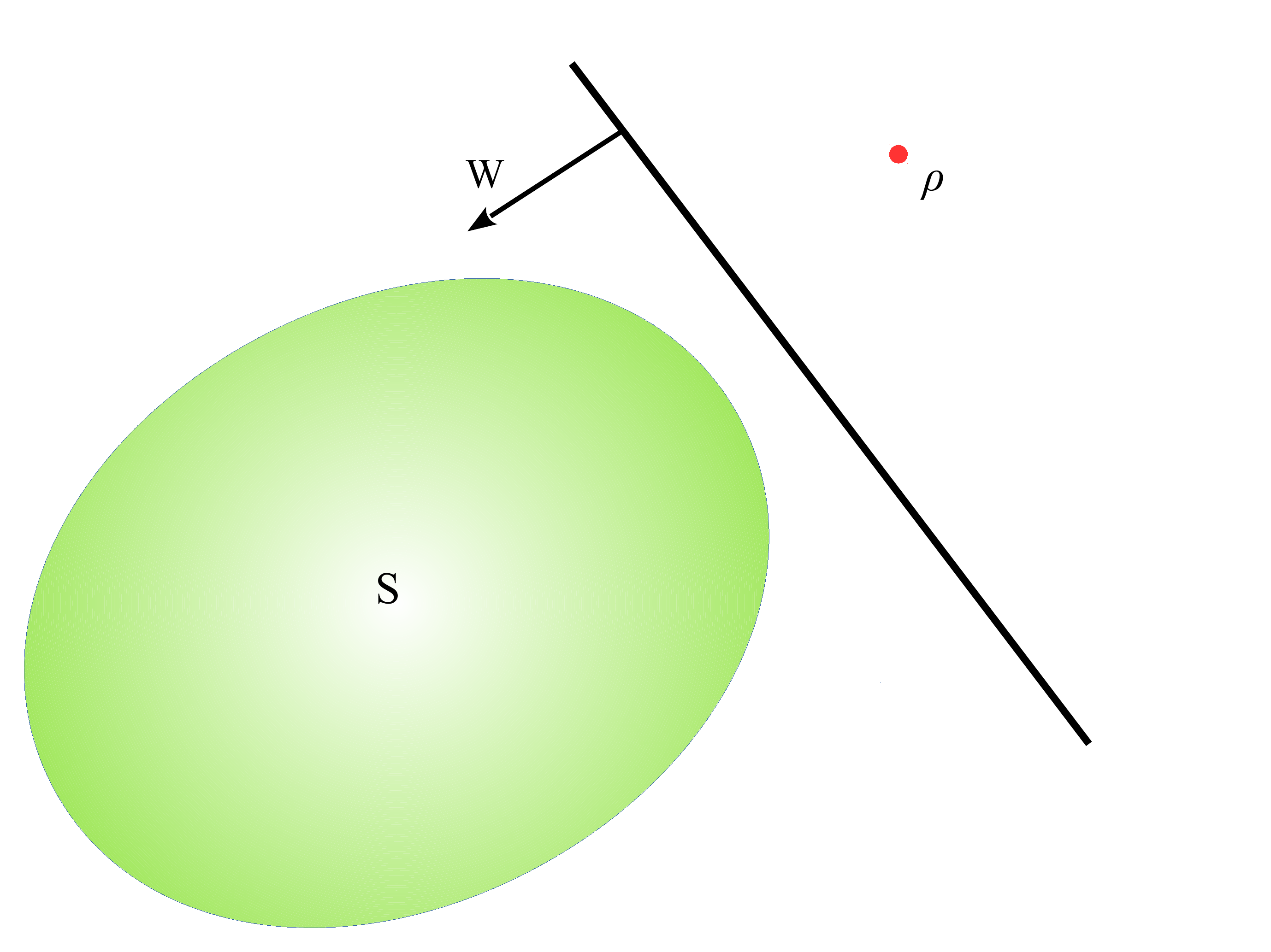}
\end{center}
\caption[Schematic picture of the Hahn-Banach theorem]
{Schematic picture of the Hahn-Banach theorem. The (unique)
unit vector
orthonormal to the hyperplane can be used to define \emph{right} and
\emph{left} in respect to the hyperplane by using the signum of the scalar product.}\label{fig:hyper}
\end{figure}

The statement of the theorem is illustrated in figure \ref{fig:hyper}. The figure
motivates the introduction of a new coordinate system
located within the hyperplane (supplemented by an orthogonal vector $W$ which
is chosen such that it points away from $S$).
Using this coordinate system, every state $\rho$ can be characterized by
its distance from the plane, by projecting $\rho$ onto the chosen orthonormal
vector and using the trace as scalar product, i.e. $\mbox{tr}(W\rho)$. This
measure is either positive, zero, or negative. We now suppose that \(S\) is the convex compact set of all separable states. 
According to our choice of
basis in figure \ref{fig:hyper}, every separable state has a positive
distance while there are some entangled states with a negative distance.
More formally this can be phrased as:\\
\textbf{Def. 7}
\emph{A Hermitian operator (an observable) $W$ is called an entanglement
witness (EW)\index{Entanglement witness (EW)} if and only if}
\begin{equation}
\exists \ \rho \quad \mbox{such that} \quad \mbox{tr}(W \rho) < 0, \qquad
\mbox{while} \quad \forall \sigma \in S, \quad \mbox{tr}(W\sigma) \geq 0.
\end{equation}


\noindent \textbf{Def. 8}\index{Decomposable entanglement witness}
\emph{An EW is decomposable if and only if there exists operators $P$, $Q$ with}
\begin{equation}
W = P + Q^{T_{A}}, \qquad P,Q \geq 0.
\end{equation}

\noindent \textbf{Lemma 2}\\
\emph{Decomposable EW cannot detect PPT entangled states.}

\noindent Proof:\\
Let $\delta$ be a PPT entangled state and $W$ be a decomposable EW.  Then
\begin{equation}\label{eqn:dEWnd}
\mbox{tr}(W\delta) = \mbox{tr}(P\delta) + \mbox{tr}(Q^{T_{A}} \delta)
= \mbox{tr}(P\delta)+\mbox{tr}(Q\delta^{T_{A}}) \geq 0.
\end{equation}
Here we used Lemma 1.\(\square\)

\noindent \textbf{Def. 9}
\index{Non-decomposable entanglement witness (nd-EW)}
\emph{An EW is called non-decomposable entanglement witness (nd-EW) if and only if
there exists at least one PPT entangled state which is detected by the witness.}

Using these definitions, we can restate the consequences of the
Hahn-Banach theorem in several ways:\\
\textbf{Theorem 9} \cite{Woronowicz,HorodeckiPPT, Terhal, sakkhi-ager}
\begin{enumerate}
\item $\rho$ \emph{is entangled if and only if,  $\exists$ a witness $W$, such that} $\mbox{tr}(\rho W) < 0$.
\item $\rho$ \emph{is a PPT entangled state  if and only if  $\exists$ a nd-EW, $W$, such that} $\mbox{tr}(\rho W) < 0$.
\item $\sigma$ \emph{is separable  if and only if  $\forall$ EW,} $\mbox{tr}(W\sigma) \geq 0$.
\end{enumerate}
From a theoretical point of view, the theorem is quite powerful.
However, it does not give any insight of how to construct for a given state $\rho$, the appropriate witness operator.

\begin{figure}
\begin{center}
\includegraphics[scale=0.3]{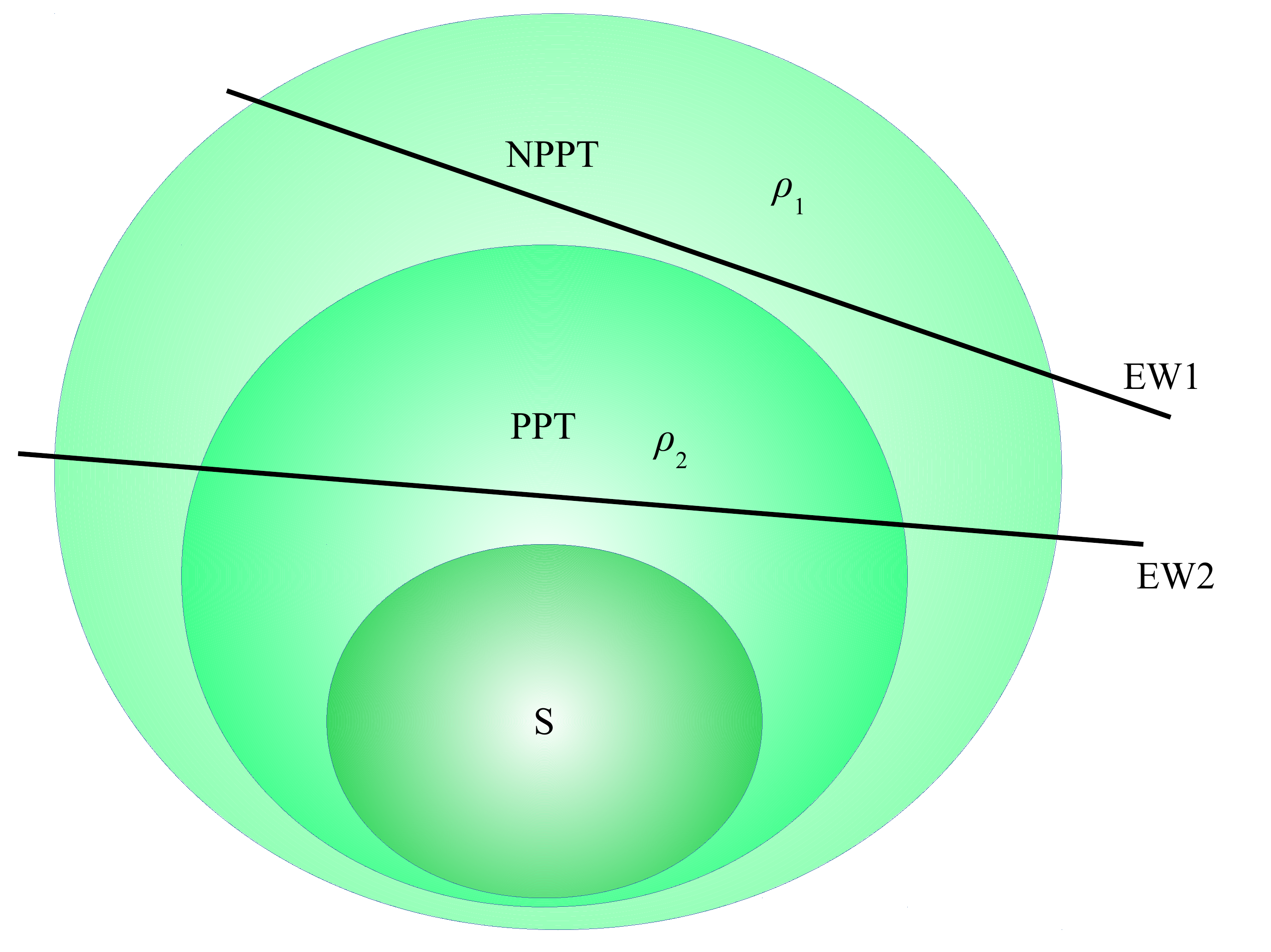}
\end{center}
\caption[Comparison of a non-decomposable entanglement witness with a decomposable one]
{Schematic view of the Hilbert-space with two states $\rho_1$ and
$\rho_2$ and two witnesses $EW1$ and $EW2$. $EW1$ is a decomposable EW, and it
 detects only  NPT states like $\rho_1$. $EW2$ is an  nd-EW, and it
detects also some PPT states like $\rho_2$. Note that neither witness
detects \emph{all} entangled states.}\label{fig:schhil}
\end{figure}
\subsubsection{Examples}\label{sec:E}
For a  decomposable witness
\begin{equation}
W' = P + Q^{T_{A}},
\end{equation}
\begin{equation}
\label{narayani}
\mbox{tr}(W'\sigma) \geq 0,
\end{equation}
for all separable states \(\sigma\).

\noindent Proof:\\
If $\sigma$ is separable, then it can be written as a convex sum of product
vectors.
So if Eq. (\ref{narayani}) holds for any product vector \(|e,f\rangle\),
any separable state will  also satisfy the same.
\begin{eqnarray}
\mbox{tr}(W' |e,f\rangle \langle e,f|) &=&
\langle e,f |W' |e,f\rangle \nonumber \\
&=&
\underbrace{\langle e,f | P |e,f \rangle}_{\geq 0} + \underbrace{\langle e,f | Q^{T_{A}} |e,f \rangle }_{\geq 0} , 
\end{eqnarray}
because
\begin{equation}
\langle e,f | Q^{T_{A}} |e,f \rangle = \mbox{tr} (Q^{T_{A}} |e,f\rangle \langle e,f |) = \mbox{tr} (Q |e^\ast,f \rangle \langle e^\ast,f |  ) 
\geq 0.
\end{equation}
Here we used Lemma 1, and $P,Q \geq 0$. \(\square\)

This argumentation shows that $W=Q^{T_{A}}$ is a suitable witness also.
Let us now consider 
the simplest case of ${\mathbb{C}}^2 \otimes {\mathbb{C}}^2$.  We can use
\begin{equation}
|\Phi^+ \rangle = \frac{1}{\sqrt{2}}\left(|00\rangle + |11\rangle\right),
\end{equation}
to write the density matrix
\begin{equation}
Q = \left( \begin{array}{cccc}
\frac12 & 0 & 0 & \frac12 \\
0 & 0 & 0 & 0 \\
0 & 0 & 0 & 0 \\
\frac12 & 0 & 0 & \frac12 \end{array}\right).  \quad \mbox{Then} \quad
Q^{T_{A}} = \left( \begin{array}{cccc}
\frac12 & 0 & 0 & 0 \\
0 & 0 & \frac12 & 0 \\
0 & \frac12 & 0 & 0 \\
0 & 0 & 0 & \frac12 \end{array}\right).
\end{equation}
One can quickly verify that indeed $W=Q^{T_{A}}$ fulfills the witness
requirements. Using
\begin{equation}
|\Psi^- \rangle = \frac{1}{\sqrt{2}} \left(|01 \rangle - |10 \rangle\right),
\end{equation}
we can rewrite the witness:
\begin{equation}\label{eqn:exwit}
W = Q^{T_{A}} = \frac{1}{2} \left(I_4 - 2 |\Psi^- \rangle \langle \Psi^- |\right),
\end{equation}
where $I_4$ denotes the identity operator on $\mathbb{C}^2 \otimes \mathbb{C}^2$.
This witness now detects $|\Psi^- \rangle$:
\begin{equation}
\mbox{tr} (W |\Psi^- \rangle \langle \Psi^- |) = -\frac{1}{2}.
\end{equation}


\subsection{Positive maps}\index{Positive maps (PM)}

\subsubsection{Introduction and definitions}
So far we have only considered states 
belonging to a Hilbert space ${\cal H}$, and operators acting on the
Hilbert space. However, the space of operators ${\cal B}({\cal H})$ has also a Hilbert space structure. 
We now look at transformations of operators, the so-called maps\index{Map} which can be regarded as 
\emph{superoperators}\index{Super operator}. As we will see, this will lead us to an important 
characterization of entangled and separable states.
We start by defining linear maps:\\
\textbf{Def. 10}
\emph{A linear, self-adjoint map $\epsilon$ is a transformation} \index{Linear map} \index{Self-adjoint map}
\begin{eqnarray}
\epsilon: {\cal B} ({\cal H}_{B}) \rightarrow {\cal B} ({\cal H}_{C}),
\end{eqnarray}
\emph{which}
\begin{itemize}
\item \emph{is linear, i.e.}
\begin{eqnarray}
\epsilon(\alpha O_1+\beta O_2) &= \alpha \epsilon(O_1) + 
\beta \epsilon(O_2)\quad\forall O_1,\,O_2\in {\cal B}({\cal H}_{B}),
\end{eqnarray}
\emph{where \(\alpha\), \(\beta\) are complex numbers,}
\item \emph{and maps Hermitian operators onto Hermitian operators, i.e.}
\begin{eqnarray}
\epsilon(O^{\dagger})&=\left(\epsilon(O)\right)^{\dagger}\qquad\forall O \in {\cal B}({\cal H}_{B}).
\end{eqnarray}
\end{itemize}
For brevity, we will only write ``linear map'', instead of ``linear self adjoint map''. 
The following definitions help to further characterize linear maps.\\
\textbf{Def. 11} \emph{A linear map $\epsilon$ is called trace preserving if }
\begin{eqnarray}
\mbox{tr}(\epsilon(O))=\mbox{tr}(O)\quad\forall O\in {\cal B}({\cal H}_{B}).
\end{eqnarray}

\noindent \textbf{Def. 12} \emph{Positive map:}\\
\emph{A linear, self-adjoint map $\epsilon$ is called positive if}
\begin{eqnarray}
\forall \rho \in {\cal B}({\cal H}_{B}), \quad
\rho \geq 0 \quad \Rightarrow \quad
\epsilon(\rho) \geq 0.
\end{eqnarray}
Positive maps have, therefore, the property of mapping positive operators onto
positive operators. It turns out that by considering maps that are a tensor product of
a positive operator acting on subsystem A, and the identity acting  on subsystem
B, one can learn about the properties of the composite system.  

\noindent \textbf{Def. 13} \emph{Completely positive map:} \index{Completely positive map}\\
\emph{A positive linear map $\epsilon$ is completely positive if for any tensor extension
of the form}
\begin{equation}
\epsilon'  = {\cal I}_A\otimes\epsilon,
\end{equation}
where 
\begin{equation}
\epsilon' :        {\cal B}({\cal H}_{A} \otimes {\cal H}_{B}) 
   \rightarrow  {\cal B}({\cal H}_{A} \otimes {\cal H}_{C}), 
\end{equation}
\emph{$\epsilon'$ is positive. Here \({\cal I}_A\) is the identity map on \({\cal B}({\cal H}_{A})\).}

\noindent\textbf{Example: Hamiltonian evolution of a quantum system.}
Let $O\in {\cal B}({\cal H}_B)$ and $U$ an unitary matrix and let us define $\epsilon$ by
\begin{eqnarray}
\epsilon:  {\cal B}( {\cal H}_A) & \rightarrow & {\cal B}( {\cal H}_A) \nonumber \\
 \epsilon(O) & = & UOU^{\dagger}.
\end{eqnarray}
As an example for this map, consider the time-evolution of a density matrix. It
can be written as $\rho(t)=U(t)\rho(0)U^{\dagger}(t)$, i.e. in the form given above. 
Clearly this map is linear, self-adjoint, positive
and trace-preserving. It is also completely positive, because for 
$0\leq w\in {\cal B}({\cal H}_A \otimes {\cal H}_A)$,
\begin{equation}
({\cal I}_A\otimes\epsilon)w  = (I_A\otimes U)w(I_A\otimes U^{\dagger})
=\tilde{U}w\tilde{U}^{\dagger},
\end{equation}
where $\tilde{U}$ is unitary. But then $\langle \psi |\tilde{U}w\tilde{U}^{\dagger} |\psi \rangle \geq 0$,
if and only if $\langle \psi | w |\psi \rangle \geq 0$ (since positivity is not changed by unitary evolution).

\noindent\textbf{Example: Transposition.}
An example of a positive but not completely positive map is the transposition $T$ defined as:
\begin{eqnarray}
T:  {\cal B}({\cal H}_B) & \rightarrow & {\cal B}({\cal H}_B) \nonumber \\
 T(\rho) & = & \rho^{T}.
\end{eqnarray}
Of course this map is positive, but it is not completely positive, because
\begin{equation}
({\cal I}_A\otimes T)w=w^{T_{B}},
\end{equation}
and we know that there exist states for which $\rho\geq 0$, but $\rho^{T_{B}}\not\geq 0$.

\noindent \textbf{Def. 14}
\emph{A positive map is called decomposable if and only if it can be written as}
\begin{equation}
\epsilon=\epsilon_1+\epsilon_2 T
\end{equation}
\emph{where $\epsilon_1$, $\epsilon_2$ are completely positive maps and $T$ is the 
operation of transposition.}

\subsubsection{Positive maps and entangled states}

Partial transposition can be regarded as a particular case of a map that is positive but not completely positive. 
We have already seen that this particular positive but not completely positive map gives us a 
way to discriminate entangled states from separable states.  
The theory of positive maps provides with stronger conditions for separability, as shown in Ref. \cite{HorodeckiPPT}.\\
\textbf{Theorem 10}\\
\emph{A state $\rho\in {\cal B} ({\cal H}_A \otimes {\cal H}_B)$ is separable if and only if for all positive maps}
\begin{equation}
\epsilon:  {\cal B} ({\cal H}_B)  \rightarrow {\cal B} ({\cal H}_C),
\end{equation}
\emph{we have}
\begin{equation}
({\cal I}_A\otimes\epsilon)\rho\geq 0.
\end{equation}

\noindent Proof: 

[$\Rightarrow$] As $\rho$ is separable,   we can write it as
\begin{equation}
\rho
=\sum_{k=1}^P p_k |e_k \rangle \langle e_k | \otimes |f_k \rangle \langle f_k |,
\end{equation}
for some $P>0$. On this state, $({\cal I}_A\otimes\epsilon)$ acts as
\begin{equation}
({\cal I}_A\otimes\epsilon)\rho 
=\sum_{k=1}^P p_k |e_k \rangle \langle e_k | \otimes \epsilon \left(|f_k \rangle \langle f_k |\right)
\geq 0,
\end{equation}
where the last $\geq$ follows because $|f_k \rangle \langle f_k |\geq0$, and $\epsilon$ is positive.\\
$[\Leftarrow]$ The proof in this direction is not as easy as the only if direction. We shall prove it 
at the end of this section.\\
Theorem 10 can also be recast into the following form:\\
\textbf{Theorem 11} \cite{HorodeckiPPT}\\
\emph{A state $\rho\in {\cal B} ({\cal H}_A \otimes {\cal H}_B)$  is entangled if and only if there exists a positive
map \(\epsilon:  {\cal B} ({\cal H}_B)  \rightarrow {\cal B} ({\cal H}_C)\), such that}
\begin{equation}
\label{chot-bogoley}
({\cal I}_A\otimes\epsilon)\rho\not\geq 0.
\end{equation}
Note that Eq. (\ref{chot-bogoley}) can never hold for maps, \(\epsilon\), that are completely positive, and for 
non-positive maps, it may hold even for separable states. Hence, any positive but not completely positive map 
can be used to detect entanglement. 

\subsubsection{Choi-Jamio{\l}kowski Isomorphism}
In order to complete the proof of Theorem 10,
we introduce first the 
Choi-Jamio{\l}kowski isomorphism \cite{jamiolkowski}
\index{Choi-Jamio{\l}kowski isomorphism}
between operators and maps.
Given an operator $E\in {\cal B} ({\cal H}_B \otimes {\cal H}_C)$, and an orthonormal product basis $|k,l \rangle$,
we define a map by
\begin{eqnarray}
\epsilon: {\cal B} ({\cal H}_B)  &\rightarrow & {\cal B} ({\cal H}_C) \nonumber \\
\epsilon(\rho) & = & \sum_{k_1,l_1,k_2,l_2}\,{}_{BC}\langle k_1l_1| E | k_2l_2 \rangle_{BC} \quad
|l_1\rangle_{CB} \langle k_1 | \rho |k_2\rangle_{BC} \langle l_2|,
\end{eqnarray}
or in short form,
\begin{equation}
\epsilon(\rho)= \mbox{tr}_{B}(E\rho^{T_{B}}).
\end{equation}
This shows how to construct the map $\epsilon$ from a given operator $E$. To construct an
operator from a given map we use the state
\begin{equation}
|\Psi^+ \rangle =\frac{1}{\sqrt{M}}\sum_{i=1}^{M} |i\rangle_{B'} |i\rangle_{B}
\end{equation}
(where $M=\dim {\cal H}_B$) to get
\begin{equation}
M\left(I_{B'}\otimes\epsilon\right)\left( |\Psi^+\rangle \langle \Psi^+| \right)=E.
\end{equation}
This isomorphism between maps and operators results in the following properties:\\
\textbf{Theorem 12} \cite{jamiolkowski, Woronowicz,HorodeckiPPT, Terhal, sakkhi-ager}
\emph{
\begin{enumerate}
\item $E\geq0$ if and only if $\epsilon$ is a completely positive map.
\item $E$ is an entanglement witness if and only if $\epsilon$ is a positive map.
\item $E$ is a decomposable entanglement witness if and only if $\epsilon$ is decomposable.
\item $E$ is a non-decomposable entanglement witness if and only if $\epsilon$ is non-decomposable and positive.
\end{enumerate}
}
To indicate further how this equivalence between maps and operators works, we develop here a proof for  the ``only if" direction of the second statement. Let
$E \in {\cal B}({\cal H}_B \otimes {\cal H}_C)$ be an entanglement witness, then $\langle e,f| E |e,f \rangle \geq 0$.
By the Jamio{\l}kowski isomorphism, the corresponding map is defined as 
$\epsilon(\rho)=\mbox{tr}_{B}(E\rho^{T_{B}})$ where $\rho \in {\cal B}({\cal H}_B)$.
We have to show that
\begin{equation}
{}_C \langle \phi | \epsilon(\rho) |\phi\rangle_C={}_C \langle \phi| \mbox{tr}(E\rho^{T_{B}}) |\phi \rangle_C \geq 0\qquad 
\forall |\phi\rangle_C \in {\cal H}_C.
\end{equation}
Since $\rho$ acts on Bob's space, using the spectral decomposition of $\rho$,
\(\rho=\sum_i\lambda_i |\psi_i \rangle \langle \psi_i| \), leads to 
\begin{equation}
\rho^{T_{B}}=\sum_i\lambda_i 
|\psi_i^\ast \rangle \langle \psi_i^\ast |,
\end{equation}  
where all $\lambda_i\ge 0$. Then
\begin{eqnarray}
{}_C\langle\phi | \epsilon(\rho)|\phi\rangle_C  &=&
{}_C\langle\phi | \sum_i \mbox{tr}_B (E\lambda_i |\psi_i^\ast\rangle_B {}_B \langle \psi_i^\ast |) |\phi\rangle_C \nonumber\\
&=& \sum_i\lambda_i {}_{BC} \langle \psi_i^\ast,\phi | E |\psi_i^\ast,\phi \rangle_{BC} \geq 0.
\end{eqnarray}
\(\square\)

We can now proof the $\Leftarrow$ direction of Theorem 10
or, equivalently,
the $\Rightarrow$ direction of Theorem 11.
We thus have to show that if $\rho_{AB}$
is entangled, there exists a positive map $\epsilon: {\cal B}({\cal H}_A) \rightarrow {\cal B}({\cal H}_C)$,
such that $\left(\epsilon\otimes {\cal I}_B\right)\rho$ is not positive definite.
If $\rho$ is entangled, then there exists an entanglement witness
$W_{AB}$ such that
\begin{eqnarray}
\mbox{tr}(W_{AB}\rho_{AB}) < 0, \quad \mbox{and} \nonumber \\
\mbox{tr}(W_{AB}\sigma_{AB}) \geq 0,
\end{eqnarray}
for all separable $\sigma_{AB}$. $W_{AB}$ is an entanglement witness (which detects
$\rho_{AB}$) if and only if $W_{AB}^{T}$ (note the complete transposition!) is also an entanglement witness (which detects
$\rho_{AB}^{T}$).
%
We define a map by
\begin{eqnarray}
\epsilon: {\cal B}({\cal H}_A) & \rightarrow & {\cal B}({\cal H}_C),\\
 \epsilon(\rho) & = & \mbox{tr}_{A} (W^{T}_{AC}\rho^{T_{A}}_{AB}),
\end{eqnarray}
where $\dim {\cal H}_C=\dim {\cal H}_B = M$. Then
\begin{equation}
(\epsilon\otimes {\cal I}_B)(\rho_{AB})= \mbox{tr}_{A}(W_{AC}^{T}\rho_{AB}^{T_{A}})=\mbox{tr}_{A}(W_{AC}^{T_{C}}\rho_{AB})=
\tilde{\rho}_{CB},
\end{equation}
where we have used Lemma 1,
and that $T=T_{A} \circ T_{C}$.
To complete the proof, one has to show that  $\tilde{\rho}_{CB}\not\ge 0$, which can be done by showing that 
\({}_{CB}\langle \psi^+| \tilde{\rho}_{CB} |\psi^+\rangle_{CB} < 0\), where 
\(|\psi^+ \rangle_{CB}=\frac{1}{\sqrt{M}} \sum_i |ii\rangle_{CB}\), with \(\{|i\rangle\}\) being  an orthonormal basis. 
\(\square\)

\subsection{Range criterion} \index{Range criterion}
The range criterion \cite{Pawelbound} gives a non-operational condition for separability, which is based on the range of the density matrix and is, in particular, independent of the partial transposition criterion. The range criterion may not detect inseparability in some states for which the partial transposition criterion succeeds but it works efficiently in many cases, especially for the bound entangled states, where the other one fails.

\noindent\textbf{Def. 15} \index{Range of a matrix}\emph{Range of a matrix $M$ on a Hilbert space $\mathcal{H}$ is defined as the span (i.e., the set of all possible linear combinations) of its column vectors. Alternatively, it can be defined as $\mathcal{R}(M) \equiv \{\Ket{\psi} \in \mathcal{H} \ | \ M\Ket{\phi}= \Ket{\psi} \mbox{ for some } \Ket{\phi} \in \mathcal{H}\}$.}

It can be easily shown that for a density matrix $\rho$ in a Hilbert space $\mathcal{H}$ having spectral decomposition 
\begin{equation}
\rho = \sum_i p_i \Ket{\psi_i}\Bra{\psi_i},
\end{equation}where $p_i > 0, \forall i$, the set of vectors $\{\Ket{\psi_i}\}$ spans the range of $\rho$, $\mathcal{R}(\rho)$. The range criteria of separability is given by the following theorem.

\noindent\textbf{Theorem 13} \cite{Pawelbound} \\
\emph{Let $\rho_{AB}$ be a state on the Hilbert space $\mathcal{H} = \mathcal{H}_A \otimes \mathcal{H}_B$, where $d_A$ and $d_B$ are the dimensions of $\mathcal{H}_A$ and $\mathcal{H}_B$ respectively. If $\rho_{AB}$ is separable, then there exists a set of product vectors of the form $\Ket{\psi_{i}} \otimes \Ket{\phi_{k}}$ and non-zero probabilities $ p_{ik}$, where $\{i,k\}$ belongs to a set of  $N \leq d_A^2 d_B^2$ pairs of indices, such that the following conditions hold:
\begin{enumerate}
\item The ensembles $\{p_{ik},\Ket{\psi_{i}} \otimes \Ket{\phi_{k}}\}$,   and $\{p_{ik},\Ket{\psi_{i}} \otimes \Ket{\phi^*_{k}}\}$ correspond to the states $ \rho_{AB} $ and $ \rho_{AB}^{T_{B}} $ respectively, i.e.,  $\rho_{AB} =\sum_{i,k} p_{ik} \left(\Ket{\psi_{i}} \otimes \Ket{\phi_{k}}\right) \left(\Bra{\psi_{i}} \otimes \Bra{\phi_{k}}\right)$ and $\rho_{AB}^{T_B} =\sum_{i,k} p_{ik} \left(\Ket{\psi_{i}} \otimes \Ket{\phi^*_{k}}\right) \left(\Bra{\psi_{i}} \otimes \Bra{\phi^*_{k}}\right)$.
\item The vectors $\{\Ket{\psi_{i}} \otimes \Ket{\phi_{k}}\}$ and $\{\Ket{\psi_{i}} \otimes \Ket{\phi_{k}^{*}}\}$ span the ranges of $\rho_{AB}$ and $ \rho_{AB}^{T_{B}} $ respectively.
\end{enumerate}
Otherwise, the state $\rho_{AB}$ must be entangled.
} \\
Proof: \\
If $\rho_{AB}$ is a separable state on the Hilbert space $\mathcal{H} = \mathcal{H}_A \otimes \mathcal{H}_B$, it can be written as a convex combination of $N \leq d_A^2 d_B^2$ products of projectors $P_{\psi_i} \otimes Q_{\phi_k}$ as 
\begin{equation}
\rho_{AB} = \sum_{i,k} p_{ik} P_{\psi_i} \otimes Q_{\phi_k} =  \sum_{i,k} p_{ik} \left(\Ket{\psi_{i}} \otimes \Ket{\phi_{k}}\right) \left(\Bra{\psi_{i}} \otimes \Bra{\phi_{k}}\right),
\label{eq:range1}
\end{equation}
where  $p_{ik}$ are non-zero probabilities satisfying $\sum_{i,k}p_{ik} = 1$. Now the transposition of a Hermitian operator is simply equivalent to the complex conjugation of its matrix elements, i.e., $Q_{\phi_k}^T = Q_{\phi_k}^* = \Ket{\phi^*_k}\Bra{\phi^*_k}$. Thus we obtain the partial transposition of $\rho_{AB}$ as
\begin{equation}
\rho_{AB}^{T_B} =\sum_{i,k} p_{ik} P_{\psi_i} \otimes Q^T_{\phi_k} = \sum_{i,k} p_{ik} \left(\Ket{\psi_{i}} \otimes \Ket{\phi^*_{k}}\right) \left(\Bra{\psi_{i}} \otimes \Bra{\phi^*_{k}}\right).
\label{eq:range2}
\end{equation}
It is evident from Eq. (\ref{eq:range1}) and (\ref{eq:range2}), that the ensembles $\{p_{ik},\Ket{\psi_{i}} \otimes \Ket{\phi_{k}}\}$,   and $\{p_{ik},\Ket{\psi_{i}} \otimes \Ket{\phi^*_{k}}\}$ correspond to the states $ \rho_{AB} $ and $ \rho_{AB}^{T_{B}} $ respectively, and the vectors $\{\Ket{\psi_{i}} \otimes \Ket{\phi_{k}}\}$ and $\{\Ket{\psi_{i}} \otimes \Ket{\phi_{k}^{*}}\}$ span the ranges of $\rho_{AB}$ and $ \rho_{AB}^{T_{B}} $ respectively. $\square$

As an example, let us consider a state in $\mathbb{C}^3 \otimes \mathbb{C}^3$, given by
\begin{eqnarray}
\rho_{AB}(a)={1 \over 8a + 1}
\left[ \begin{array}{ccccccccc}
     a &0&0&0&a&0&0&0& a   \\
      0&a&0&0&0&0&0&0&0     \\
      0&0&a&0&0&0&0&0&0     \\
      0&0&0&a&0&0&0&0&0     \\
     a &0&0&0&a&0&0&0& a     \\
      0&0&0&0&0&a&0&0&0     \\
0&0&0&0&0&0&{1+a \over 2}&0&{\sqrt{1-a^2} \over 2}\\
      0&0&0&0&0&0&0&a&0     \\
     a &0&0&0&a&0&{\sqrt{1-a^2} \over 2}&0&{1+a \over 2}\\
      \end{array}
    \right ], \ \ \
\label{tran}
\end{eqnarray}
with $0 < a < 1$. The partial transposition of this density matrix, $\rho_{AB}^{T_B}(a) $ turns out to be positive. In Ref. \cite{Pawelbound}, it was demonstrated that $\rho_{AB}(a)$ is entangled, which can be successfully detected by the range criterion. For $ a \neq 0,1$, one can find all product vectors $\{\Ket{\psi_{i}} \otimes \Ket{\phi^*_{k}}\}$ in the range of $ \rho_{AB}^{T_{B}}(a) $. It was shown that the partial complex conjugation with respect to $B$ i.e.,  $ \{\Ket{\psi_{i}} \otimes \Ket{\phi_{k}}\} $ cannot span the range of $ \rho_{AB}(a) $, thus violating the condition 2 of Theorem 13.

\section{Bell inequalities} \index{Bell inequalities}

The concept of locality with respect to shared quantum states was first brought into light by Einstein, Podolsky, and Rosen (EPR) in their seminal paper in 1935 \cite{EPRparadox}. They argued that since nonclassical correlations of entangled states of the form $|\psi_{AB}\rangle = \sum _{i} a_i |e_i\rangle \otimes |f_i\rangle$ cannot be explained by any physical theory satisfying the notions of ``locality" and ``realism", quantum mechanics must be incomplete. In 1964, J. S. Bell \cite{Bell} provided a formulation of the problem that made the assumptions of locality and realism more precise, and more importantly,
showed that the assumptions are actually \emph{testable} in experiments. He derived a mathematical inequality that must be satisfied by any physical theory of nature, which is local as well as realistic.
%
%

As we shall see, Bell inequalities are essentially a special type of entanglement witness. 
An additional property of Bell inequalities is that any entangled state detected by them is nonclassical 
in a particular way: It violates ``local realism''. 
%
%
%
%
The inequality is 
actually a constraint on a linear function of results of certain experiments.
Modulo some so-called loopholes (see e.g. \cite{gerakol}), these inequalities have been shown to be actually violated in
 experiments (see e.g. \cite{Paris-ghyama} and references therein). 
In this section, we will first derive a Bell  inequality \footnote{We do not derive here the original 
Bell inequality, which Bell derived in 1964 \cite{Bell}. Instead, we  derive the stronger form of the Bell 
inequality which Clauser, Horne,  Shimony, and Holt (CHSH) derived in 1969 \cite{CHSHineq}. 
A similar  derivation was also given by 
Bell himself in 1971 \cite{Bell71}.}
and then show how this inequality 
is violated by the singlet state.

Consider a two spin-1/2 particle state where the two particles are far apart. Let the particles
be called
\(A\) and \(B\). Let projection valued  measurements in the directions
 \(a\) and \(b\) be done on \(A\) and \(B\) respectively.  The 
outcomes of the measurements performed on the particles \(A\) and \(B\) in the 
directions \(a\) and \(b\), are respectively \(A_a\) and \(B_b\). The measurement result 
\(A_a\) (\(B_b\)), whose values can be \(\pm 1\),
 may depend on the direction \(a\) (\(b\))
  and some other uncontrolled parameter \(\lambda\) which may depend on anything, that is, 
may depend upon system or measuring device or both. Therefore we assume that \(A_a\) (\(B_b\)) 
has a definite pre-measurement value \(A_a (\lambda)\) (\(B_b (\lambda)\)). 
Measurement merely uncovers this value. This is the \emph{assumption of reality}.  \(\lambda\) is
usually called a hidden variable and this assumption is also termed as the hidden variable assumption. 
Moreover, the measurement result at $A$ ($B$) does not depend on what measurements are performed at $B$ ($A$). That is,
for example \(A_a(\lambda)\) does not depend upon \(b\). This is the \emph{assumption of locality},
also called the Einstein's locality  assumption.
The parameter \(\lambda\) 
is
assumed to have 
 a probability distribution, say \(\rho(\lambda)\). Therefore
 \(\rho(\lambda)\) satisfies the following:
\begin{equation}
\int \rho(\lambda) d\lambda = 1, \quad \rho(\lambda) \geq 0.
\end{equation}
The correlation  function of the two spin-1/2 particle state 
for a measurement in a fixed direction \(a\) for particle \(A\) and 
\(b\) for particle \(B\), is then given by (provided the hidden variables exist)
\begin{equation}
\label{eq_correlation}
E(a, b) = \int A_a(\lambda) B_b(\lambda) \rho(\lambda) d\lambda.
\end{equation}
Here 
\begin{equation}
A_a(\lambda) = \pm 1, \quad  \mbox{and} \quad  B_b(\lambda) = \pm 1,
\end{equation}
because the measurement values were assumed to be \(\pm 1\).

Let us now suppose that the observers at the  
two particles \(A\) and \(B\) can choose their measurements 
from two observables \(a\), \(a{'}\) and \(b\), \(b{'}\) respectively,
and the corresponding 
outcomes are \(A_a\), \(A_{a{'}}\) and \(B_b\), \(B_{b{'}}\) respectively. Then
\begin{eqnarray}
E(a, b) + E(a, b{'}) + E(a{'}, b) - E(a{'}, b{'}) \nonumber \\ = 
\int [A_a (\lambda)(B_b(\lambda)+ B_{b{'}}( \lambda)) + 
       A_{a{'}} (\lambda)(B_b(\lambda)- B_{b{'}}( \lambda)) ]
                \rho(\lambda) d\lambda .
\end{eqnarray}
Now \(B_b(\lambda)+ B_{b{'}}( \lambda)\)  and \(B_b(\lambda)- B_{b{'}}( \lambda)\) can only
be \(\pm 2\) and \(0\), or  \(0\) and \(\pm 2\) respectively. Consequently,
\begin{equation}
\label{eq_Bell_inequality}
-2 \leq E(a, b) + E(a, b{'}) +  E(a{'}, b) - E(a{'}, b{'})  \leq 2.
\end{equation}
This is the well-known CHSH inequality.  Note here that in obtaining the above inequality,
we have never used quantum mechanics. We have only assumed Einstein's locality principle and
an underlying hidden variable model.
Consequently, a Bell inequality is a constraint that any physical theory that is both, local and realistic, has to
satisfy. 
Below,  we will show that this inequality can be violated
by a quantum state.  Hence quantum mechanics is incompatible with an underlying local realistic model.

\subsection{Detection of entanglement by Bell inequality}
Let us now show how the singlet state can be detected by a Bell inequality. 
This additionally will indicate that quantum theory is incompatible with local realism.
For the singlet state $\Ket{\Psi^-} = \frac{1}{\sqrt{2}} (\Ket{01} - \Ket{10})$, the quantum mechanical prediction of the 
correlation function \(E(a, b)\) is given by 
 \begin{equation}
 \label{pukurchuri}
E(a, b) = \left \langle \Psi^{-} | \sigma_a \cdot \sigma_b| \Psi^{-} \right\rangle = - \cos(\theta_{ab}),
\end{equation} 
where \(\sigma_a = \vec{\sigma} \cdot \vec{a} \) and similarly for \(\sigma_b\). 
\(\vec{\sigma}= (\sigma_x, \sigma_y, \sigma_z)\), where \(\sigma_x\), \(\sigma_y\) and  \(\sigma_z\) are
the Pauli spin matrices.
And  \(\theta_{ab}\) is the angle 
between the two measurement directions \(a\) and \(b\).

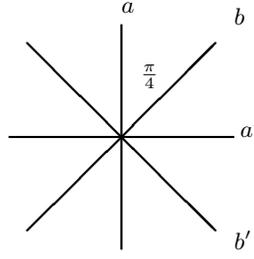
\begin{figure}[ht]
\begin{center}
\unitlength=0.5mm
\begin{picture}(80,50)(0,-10)
\thicklines
\put(10,15){\line(1,0){60}}
\put(40,48){\(a\)}
\put(70,45){\(b\)}

\put(70,15){\ \(a{'}\)}
\put(70,-15){\(b{'}\)}

\put(40,-15){\line(0,1){60}}
\put(15,-10){\line(1,1){50}}
\put(15,40){\line(1,-1){50}}

\put(45, 30){\(\frac{\pi}{4}\)}

\end{picture}
\end{center}
\caption[Optimal apparatus setting for violation of Bell-CHSH inequality by singlet.]{Schematic diagram 
showing the direction of \(a\), \(b\), \(a{'}\), \(b{'}\)
for obtaining maximal violation of Bell inequality by the singlet state.}
\label{fig_direction_highest} 
\end{figure}

So for the singlet state, one has 
\begin{eqnarray}
\label{khalchuri}
B_{CHSH}&= & E(a, b) + E(a, b{'}) +  E(a{'}, b) - E(a{'}, b{'}) \nonumber \\
&= & - \cos \theta_{ab} - \cos \theta_{a b{'}} - \cos \theta_{a{'}b} + \cos \theta_{a{'}b{'}}.
\end{eqnarray}
The maximum value of this function is attained for the directions \(a\), \(b\), \(a{'}\), \(b{'}\)
on a plane, as given in Fig. \ref{fig_direction_highest}, and in that case 
\begin{equation}
|B_{CHSH}| = 2 \sqrt{2}.
\end{equation}

This clearly violates the inequality in Eq. (\ref{eq_Bell_inequality}). But Eq. 
(\ref{eq_Bell_inequality})
was a constraint 
for any physical theory which 
has an underlying local hidden variable model.  As the singlet state,
a state allowed by the quantum mechanical description of nature, violates the constraint (\ref{eq_Bell_inequality}),
quantum mechanics cannot have an underlying local hidden variable model. In other words, quantum mechanics is not 
local realistic. This is the statement of the celebrated Bell theorem.

Moreover, it is easy to convince oneself that any separable state does have a local realistic description, so that 
such a state cannot violate a Bell inequality. Consequently, the violation of Bell inequality by the singlet state indicates 
that the singlet state is an entangled state. Further, the operator (cf. Eqs. (\ref{pukurchuri}) and (\ref{khalchuri})) 
\begin{equation}
\tilde{B}_{CHSH} = \sigma_a \cdot \sigma_b + \sigma_a \cdot \sigma_{b{'}} + \sigma_{a{'}} \cdot \sigma_b 
- \sigma_{a{'}} \cdot \sigma_{b{'}} 
\end{equation}
can, by suitable scaling and change of origin, be considered as an entanglement witness for the singlet state, for 
\(a\), \(b\), \(a{'}\), \(b{'}\) chosen as in 
figure \ref{fig_direction_highest} (cf. \cite{POMD}). 

Note that violation of Bell inequalities is stronger than entanglement. For example, the Werner state \(\rho^W_{AB}(p) = p|\Psi^-\rangle \langle \Psi^- | + \frac{1-p}{4} I \otimes I\) is  entangled for $p>1/3$, but it violates CHSH inequality for $p > 1/\sqrt{2}$ \cite{Werner, horo_bell}.


\section{Quantification of entanglement}\index{Quantification of entanglement} \index{Entanglement measures}
\label{sec:ent_measures}

The entanglement content of a pure two-party quantum state was initially quantified by the usefulness of the state in communication protocols, e.g., quantum teleportation, quantum dense coding, etc. \cite{bennett_eof0, bennett_eof1}. 
Since entangled quantum states can be used to perform teleportation and dense coding with efficiencies exceeding those situations in which no entanglement is available, entanglement is considered to be a ``resource". Moreover, it was found that the singlet state $\Ket{\Psi^-} = \frac{1}{\sqrt{2}} (\Ket{01} - \Ket{10})$ can perform these tasks with the maximal possible efficiency, thus it was assumed that the singlet state or any other state which is connected to the singlet state by local unitary transformations is a maximally entangled state in $\mathbb{C}^2 \otimes \mathbb{C}^2$. It was further assumed that maximally entangled states in $\mathbb{C}^2 \otimes \mathbb{C}^2$ has a unit amount of entanglement, or has 1 \emph{ebit} (``entanglement bit"). What if
one has a shared entangled state $\Ket{\psi_{AB}} \neq \Ket{\Psi^-}$? In that case, one can show that given many copies of $\Ket{\psi_{AB}} \in \mathbb{C}^2 \otimes \mathbb{C}^2$,  one can extract a fewer number of singlets using LOCC, which can thereafter be used in quantum communication schemes. Conversely, if one has a collection of singlets, then it can be converted into a collection of $\Ket{\psi_{AB}}$ via LOCC. Bennett et al. showed that $n$ copies of an entangled state $\Ket{\psi_{AB}}$, shared between Alice and Bob can be \emph{reversibly} converted, using only LOCC between Alice and Bob, into $m$ copies of singlets, where $m/n$ tends to $S(\rho_A) = S(\rho_B)$ and the fidelity of the conversion approaches unity for large $n$ \cite{bennett_eof1}. 
This led to the quantification of the entanglement content of a pure quantum state $\Ket{\psi_{AB}}$ by the von Neumann entropy of its reduced density matrices \cite{bennett_eof1}: \index{Entropy of entanglement} 
\begin{equation} 
\mathcal{E}_E(\Ket{\psi_{AB}}) = S(\rho_A) = S(\rho_B).
\label{ent_entropy}
\end{equation}
This quantification also remains valid in higher dimensions. We refer to the quantity $\mathcal{E}_E(\Ket{\psi_{AB}})$ as the the ``entropy of entanglement" (or simply entanglement) of $\Ket{\psi_{AB}}$. 
Clearly, for a disentangled pure state $\Ket{\psi_{AB}} = \Ket{\psi_A} \otimes \Ket{\psi_B}$, $\rho_A$ and $\rho_B$ are also pure states, for which the von Neumann entropies vanish, and $\mathcal{E}_E(\Ket{\psi_{AB}}) = 0$. But if $\Ket{\psi_{AB}}$ is an entangled state of the form $|\psi_{AB}\rangle = \sum_i a_i |e_i\rangle \otimes |f_i\rangle$ with more than one nonzero $a_i$, then we have $\rho_A = \sum_i |a_i|^2 \Ket{e_i}\Bra{e_i}$ and
$\rho_B = \sum_i |a_i|^2 \Ket{f_i}\Bra{f_i}$. In this case, we have $\mathcal{E}_E(\Ket{\psi_{AB}}) = S(\rho_A) = S(\rho_B) > 0$, and is given by the Shannon entropy of the probability distribution $\{|a_i^2|\}$. Entropy of entanglement ranges from zero for a product state to $\log_2 d$ for a maximally entangled state in a Hilbert space of dimension $d \otimes d$. Clearly for the singlet state $\Ket{\Psi^-}$, the entropy of entanglement $\mathcal{E}_E = 1$. 

%

Before extending the quantitative theory of entanglement to the more general situation in which Alice and Bob share a mixed state $\rho_{AB}$, we present essential conditions that any measure of entanglement $\mathcal{E}(\rho)$ has to satisfy \cite{vedral_ent, horo_review, horo_ent_limit}.
\begin{enumerate}
\item Non-negativity, i.e.,  $\mathcal{E}(\rho) \geq 0$ for any bipartite quantum state $\rho$.
\item Entanglement vanishes for separable states, i.e., $\mathcal{E}(\sigma) = 0$ if $\sigma$ is separable.
\item Invariance under local unitary transformations, i.e., $\mathcal{E}(U_A \otimes U_B \rho_{AB} U^{\dagger}_A \otimes U^{\dagger}_B) = \mathcal{E}(\rho_{AB})$.
\item Entanglement cannot increase under local operations and classical communication, i.e., for a given LOCC $\Lambda$, $\mathcal{E}(\Lambda(\rho)) \leq \mathcal{E}(\rho)$. 
\end{enumerate}
Condition 1 is there by convention. In any resource theory, the quantification of the resource must be done by a quantity that does not increase under the free operations. Moreover, the quantity must be zero for the states that can be created by these free operations. In the resource theory of entanglement, the free operations are the LOCC, and thus entanglement measures cannot increase under LOCC and separable states must not have any entanglement. This accounts for conditions 2 and 4.
Condition 3 arises as local unitary transformations represent only a local change of basis and do not change any correlation. A quantity that satisfies these conditions can be called an entanglement measure, and is eligible for the quantification of the entanglement content of a quantum state. They are also often referred to as \emph{entanglement monotones}.
Some authors also impose \emph{convexity} and \emph{additivity} properties for entanglement measures:
\begin{itemize}
\item Entanglement is a convex function, i.e., $\mathcal{E}\left(\sum_i p_i \rho_i\right) \leq \sum_i p_i \mathcal{E}(\rho_i)$, for an ensemble $\{p_i, \rho_i\}$.
\item Entanglement is additive, i.e., $\mathcal{E}(\rho^{\otimes n}) = n \mathcal{E}(\rho)$.
\end{itemize}
Below, we briefly discuss a few measures of entanglement. 

\subsection{Entanglement of formation} \index{Entanglement of formation}
One way to widen the theory of entanglement measures to the mixed state regime is by the convex roof extension of pure state entanglement measures \cite{uhlmann_convex}. The first measure introduced by this technique was the \emph{entanglement of formation} \cite{bennett_eof2}. \\
\noindent \textbf{Def. 16} \emph{Entanglement of formation of a  quantum state $\rho_{AB}$ shared between Alice and Bob is defined as}
\begin{equation}
\mathcal{E}_{EoF}(\rho_{AB}) = \underset{\{p_i,\Ket{\psi^i_{AB}}\}}{\mbox{min}} \sum_i p_i \mathcal{E}_E(\Ket{\psi^i_{AB}}),
\label{eq:eof}
\end{equation}
\emph{where the minimization is taken over all possible pure state decompositions,  
$\rho_{AB} = \sum_i p_i \Ket{\psi^i_{AB}}\Bra{\psi^i_{AB}}$, of $\rho_{AB}$, and $\mathcal{E}_E(\Ket{\psi^i_{AB}})$ is the entropy of entanglement of the pure state $\Ket{\psi^i_{AB}}$.} 

Clearly, the entanglement of formation for a pure state collapses to the corresponding entropy of entanglement. Using the singlet as the basic unit of entanglement, one can perceive the operational meaning of the entanglement of formation in the following manner:
\begin{enumerate}
\item Decompose $\rho_{AB}$ into a pure state ensemble as $\rho_{AB} = \sum_i p_i \Ket{\psi^i_{AB}}\Bra{\psi^i_{AB}}$.
\item Choose the state $\Ket{\psi^i_{AB}}$ according to the corresponding probability $p_i$.
\item Prepare the state $\Ket{\psi^i_{AB}}$ from singlets via local operations and classical communication.
\item Finally, forget the identity of the chosen ensemble state.
\end{enumerate}
In this way, one needs on average $\sum_i p_i \mathcal{E}_E(\Ket{\psi^i_{AB}})$ singlets, and then one can choose the pure state ensemble for which the average is minimum. This minimum number of singlets required to prepare $\rho_{AB}$ in this procedure gives the entanglement of formation of $\rho_{AB}$.


The convex roof optimization given in Eq. (\ref{eq:eof}) is formidable to compute for general mixed states. However, the exact closed form of entanglement of formation is known for two-qubit mixed states in terms of the ``\emph{concurrence}".

\subsubsection{Concurrence} \index{Concurrence}
Concurrence for pure states was first introduced in Ref. \cite{hill_conc}. For a two-qubit pure state, $\Ket{\psi_{AB}}$, the concurrence is defined as
\begin{equation}
\mathcal{C}(\psi_{AB}) = |\langle\psi_{AB}|\tilde{\psi}_{AB}\rangle|,
\label{eq:pure_conc}
\end{equation}
where $\Ket{\tilde{\psi}_{AB}} =  \sigma_y \otimes \sigma_y \Ket{\psi^*_{AB}}$, with $\Ket{\psi^*_{AB}}$ being the complex conjugate of $\Ket{\psi_{AB}}$ in the standard computational basis $\{ \Ket{00}, \Ket{01}, \Ket{10}, \Ket{11} \}$. For two-qubit mixed states, a closed form expression of the convex roof extension of concurrence can be obtained \cite{wootters_conc}. For a two-qubit density matrix $\rho_{AB}$, let us first define the spin-flipped density matrix as $\tilde{\rho}_{AB} = (\sigma_y \otimes \sigma_y) \rho^*_{AB}(\sigma_y \otimes \sigma_y)$, and the operator $R = \sqrt{\sqrt{\rho_{AB}}\rho^*_{AB}\sqrt{\rho_{AB}}}$. The convex roof extended concurrence of $\rho_{AB}$ is then given by 
\begin{equation}
\mathcal{C}(\rho_{AB}) = \mbox{min}\{0,\lambda_1 -\lambda_2 - \lambda_3 -\lambda_4\},
\label{eq:mixed_conc}
\end{equation}
where the $\lambda_i$'s are the eigenvalues of $R$ in decreasing order. A computable formula for entanglement of formation of two-qubit quantum states can be expressed in terms of the concurrence \cite{hill_conc, wootters_conc}. \\
\textbf{Theorem 14} \cite{wootters_conc} \\
\emph{The entanglement of formation of a two-qubit quantum state $\rho_{AB}$ is given by 
$\mathcal{E}_{EoF}(\rho_{AB}) = \mathcal{F}(\mathcal{C}(\rho_{AB}))$, where the function $\mathcal{F}(\mathcal{C})$ is defined as 
\begin{eqnarray}
\mathcal{F}(\mathcal{C}) = h \left( \frac{1+ \sqrt{1-\mathcal{C}^2}}{2}\right),
\end{eqnarray}
with $h(x) = -x\log_2x - (1-x)\log_2(1-x).$}

Since $\mathcal{F}(\mathcal{C})$ is a monotonically increasing function of $\mathcal{C}$ and goes from 0 to 1 as $\mathcal{C}$ goes from 0 to 1, we can also consider the concurrence as a measure of entanglement in $\mathbb{C}^2 \otimes \mathbb{C}^2$.

\subsubsection{Entanglement cost} \index{Entanglement cost}
As we have seen earlier, the entropy of entanglement of a pure state $\Ket{\psi_{AB}}$ quantifies the average number of singlets needed to asymptotically construct $\Ket{\psi_{AB}}$ via LOCC.
We went over to the mixed state scenario by using the concept of entanglement of formation. However, the definition of entanglement of formation consists of a combination of asymptotic and non-asymptotic LOCC transformations. Let us now present a purely asymptotic entanglement measure, known as \emph{entanglement cost}. \\
\noindent \textbf{Def. 17} \emph{Given $m$ copies of singlet state $(\Ket{\Psi^-}^{\otimes m})$, 
consider all LOCC protocols $\mathcal{P}$ that can transform the $m$ singlets into a state $\sigma_n$ such that $\mathcal{D}(\rho_{AB}^{\otimes n}, \sigma_n) \rightarrow 0$ as $n \rightarrow \infty$, with $\mathcal{D}$ being a suitable distance functional. The entanglement cost of $\rho_{AB}$ is defined as }
\begin{equation}
\mathcal{E}_C(\rho_{AB}) = \underset{\mathcal{P}} {\mbox{min}} \left(\underset{n \rightarrow \infty}{\lim} \frac{m}{n}\right),
\label{eq:ent_cost}
\end{equation}
\emph{where the minimization is taken over all such LOCC protocols $\mathcal{P}$.}

Hayden et al. have shown than entanglement cost is equal to the regularized entanglement of formation \cite{hayden_cost}, given by
\begin{equation}
\mathcal{E}_C = \underset{n \rightarrow \infty} {\lim} \frac{\mathcal{E}_{EoF}(\rho_{AB}^{\otimes n})}{n}.
\end{equation}
Clearly, if entanglement of formation is additive, entanglement cost will be equal to the entanglement of formation. For pure states, entanglement cost reduces to the entropy of entanglement.

\subsection{Distillable entanglement} \index{Distillable entanglement}
Distillable entanglement \cite{bennett_eof2, bennett_dist, plenio_ent, rains} is a measure dual to entanglement cost. In case of entanglement cost, we looked at the asymptotic rate at which one can prepare the given state from maximally entangled states via LOCC, whereas in this case, we will be interested in the rate of ``distillation" of a given state into singlets, via LOCC. The formal definition of  distillable entanglement of a bipartite quantum state $\rho_{AB}$ shared between Alice and Bob is as follows: 

\noindent \textbf{Def. 18} \emph{Given $n$ copies of the shared quantum state $\rho_{AB}$, if one can prepare a state $\sigma_n$ by a LOCC protocol $\mathcal{P}$, such that $\mathcal{D}\left((\Ket{\Psi^-}\Bra{\Psi^-})^{\otimes n}, \sigma_n\right) \rightarrow 0$ as $n \rightarrow \infty$, with $\mathcal{D}$ being a suitable distance functional, then $\mathcal{P}$ is referred to as a distillation protocol.
The distillable entanglement of $\rho_{AB}$ is defined as}
\begin{equation}
\mathcal{E}_D = \underset{\mathcal{P}} {\mbox{max}} \left(\underset{n \rightarrow \infty}{\lim} \frac{m}{n}\right),
\label{eq:dist_ent}
\end{equation}
\emph{where the maximization is taken over all distillation protocols $\mathcal{P}$.} 

For pure states, optimal entanglement transformations are reversible, and thus distillable entanglement and  entanglement cost coincide and reduce to the entropy of entanglement. But in general, $\mathcal{E}_D \leq \mathcal{E}_C$. To understand this inequality, we note that if the opposite is allowed, one can get more singlets by distilling a state than the amount of singlets required to create it, leading to  a \emph{perpetuum mobile}.  Bound entangled states cannot be distilled, and so the distillable entanglements for bound entangled states are always zero, whereas since these states are entangled, their entanglements of formation are non-zero. There are examples of bound entangled states, whose entanglement costs have also been proven to be non-zero \cite{vidal_bound}, leading to irreversibility in asymptotic entanglement transformations.

\subsection{Relative entropy of entanglement} \index{Relative entropy of entanglement}
\label{sec:rel_ent}

A qualitatively different way to quantify entanglement is based on the geometry of quantum states. It is defined as the distance between an entangled state and its closest separable state \cite{GM1}. If $\mathcal{S}$ is the set of all separable states, then a distance-based measure of entanglement for a bipartite shared state $\rho_{AB}$ is given by \cite{GM1, vedral_ent, vedral_rel_ent, vedral_rmp_re} 
\begin{equation}
\mathcal{E}_\mathcal{G}(\rho_{AB}) = \underset{\sigma \in \mathcal{S}} {\mbox{min}} \ \mathcal{D}(\rho_{AB}, \sigma),
\label{eq:measure_d}
\end{equation}
where $\mathcal{D}$ is a suitably chosen distance measure. For $\mathcal{E}_\mathcal{G}(\rho_{AB})$ to be a ``good"
measure of entanglement, the distance measure $\mathcal{D}$ can be required to satisfy following properties.
\begin{enumerate}
\item $\mathcal{D}(\rho, \sigma) \geq 0$ for any two states $\rho$ and $\sigma$; equality holds iff $\rho = \sigma$.
\item Invariance under unitary operations. $\mathcal{D}(U \rho U^{\dagger}, U \sigma U^{\dagger}) = \mathcal{D}(\rho, \sigma)$.
\item $\mathcal{D}(\rho, \sigma)$ is non-increasing under every completely positive and trace preserving map $\Lambda$, i.e., 
$\mathcal{D}(\Lambda(\rho), \Lambda(\sigma)) \leq \mathcal{D}(\rho, \sigma)$.
\end{enumerate} 
The reason for the distance measure $\mathcal{}$ to satisfy these properties is that they imply conditions 1 - 4 for entanglement measures mentioned earlier in this section.

One of the most famous members of this family of distance-based measures is the \emph{relative entropy of entanglement}, where we take the  von Neumann relative entropy, which is defined in analogy with the classical Kullback-Leibler distance, as the diatance measure. For two density matrices, $\rho$ and $\sigma$, it is defined as  \cite{NChuang}
\begin{equation}
S(\rho || \sigma) = \mbox{tr} \left(\rho \log_2 \frac{\rho}{\sigma}\right) = \mbox{tr} \left\{\rho (\log_2\rho - \log_2\sigma) \right\}.
\end{equation}
It is to be noted that the relative entropy $S(\rho || \sigma)$ is not symmetric in its arguments, $\rho$ and $\sigma$.

\noindent \textbf{Def. 19} \emph{The relative entropy of entanglement of a bipartite state $\rho_{AB}$ shared between Alice and Bob is defined as}
\begin{equation}
\mathcal{E}_{RE}(\rho_{AB}) = \underset{\sigma \in \mathcal{S}} {\mbox{min}} \ S(\rho_{AB} || \sigma),
\label{eq:measure_d}
\end{equation}
\emph{where $\mathcal{S}$ denotes the set of all separable states.} \\
We now state two important theorems on relative entropy of entanglement.

\noindent \textbf{Theorem 15} \cite{vedral_ent, vedral_rel_ent} \\
\emph{For pure bipartite states, the relative entropy of entanglement reduces to the entropy of entanglement. } 

\noindent \textbf{Theorem 16} \cite{vedral_rel_ent} \\
\emph{The relative entropy of entanglement $\mathcal{E}_{RE}(\rho_{AB})$ provides an upper bound to the distillable entanglement $\mathcal{E}_D(\rho_{AB})$, and a lower bound to the entanglement of formation $\mathcal{E}_{EoF}(\rho_{AB})$, i.e., $\mathcal{E}_{D}(\rho_{AB})
\leq \mathcal{E}_{RE}(\rho_{AB}) \leq \mathcal{E}_{EoF}(\rho_{AB})$.}

Although computation of the relative entropy of entanglement for arbitrary mixed states is quite hard, one can characterize
the set of entangled states for all of whom a given separable state is the closest separable state, when relative entropy is
considered as the distance measure \cite{rel_ent_closed}.


Based on other distance measures, several ``geometric'' entanglement measures have also been introduced (see Sec. \ref{sec:ksep}).

\subsection{Negativity and logarithmic negativity} \index{Negativity} \index{Logarithmic negativity}
The partial transposition criterion for entanglement, mentioned in Sec. \ref{sec:PPT}, provides another quantity to quantify the entanglement content of a given quantum state. This quantity is known as the \emph{negativity} \cite{eisert_neg, vidal_neg, karol_neg, plenio_neg}, given by the absolute sum of negative eigenvalues of the partial transposed density matrix. In other words, the negativity of a shared quantum state $\rho_{AB}$ is defined as
\begin{equation}
\mathcal{N}(\rho_{AB}) = \frac{||\rho_{AB} ^{T_B}|| - 1}{2} = \frac{||\rho_{AB} ^{T_A}|| - 1}{2},
\label{eq:neg}
\end{equation}
where $||A|| = \mbox{tr} \ \sqrt{A^{\dagger}A}$ is the matrix trace norm. Although $\mathcal{N}(\rho)$ satisfies the convexity property, it is not additive. Based on negativity, one can define an additive entanglement measure, known as \emph{logarithmic negativity}, and is given by
\begin{eqnarray}
\mathcal{E}_{LN}(\rho_{AB}) &=& \log_2 ||\rho_{AB}^{T_B}||  = \log_2 ||\rho_{AB}^{T_A}|| \nonumber \\
&=& \log_2\left(2 \mathcal{N}(\rho_{AB}) + 1 \right).
\label{eq:log_neg}
\end{eqnarray}
$\mathcal{E}_{LN}(\rho_{AB})$ is a monotone under deterministic LOCC operations.
However, it fails to be a convex function. It was also shown to be an upper bound of distillable entanglement \cite{vidal_neg}. 

A major advantage of negativity and logarithmic negativity is that they are easy to compute for general, possibly mixed, quantum states of arbitrary dimensions. Clearly, for PPT bound entangled states, $\mathcal{N}(\rho_{AB})$ and $\mathcal{E}_{LN}(\rho_{AB})$ are zero and cannot be used to quantify entanglement. But in $\mathbb{C}^2 \otimes \mathbb{C}^2$ and $\mathbb{C}^2 \otimes \mathbb{C}^3$, their non-zero values are necessary and sufficient for detecting entanglement (see Theorem 3).

\section{Classification of bipartite states with respect to quantum dense coding} \index{Quantum dense coding}
Up to now, we have been interested in splitting the set of all bipartite quantum states into separable and entangled states. 
However, one of the main motivations behind the study of entangled states is that some of them can be used to perform 
certain tasks, which are not possible if one uses states without entanglement. It is, therefore, important 
to find out which entangled states are useful for a given task. We discuss here the particular example of quantum dense coding \cite{BW}.

Suppose that Alice wants to send two bits of 
classical information to Bob.  Then a general result known as the Holevo bound (to be discussed below), shows that 
Alice must send two qubits (i.e. 2 two-dimensional quantum systems) to Bob, if only a noiseless quantum channel 
is available. However, if additionally Alice and Bob have previously shared entanglement,
then Alice may have to send less than two qubits to Bob. It was shown by Bennett and Wiesner \cite{BW}, that 
by using a previously shared singlet (between Alice and Bob), Alice will be able to send two bits to Bob, by sending just a 
single qubit.

The protocol of dense coding \cite{BW} works as follows. Assume that Alice and Bob share a singlet state 
\begin{equation}
|\Psi^-\rangle = \frac{1}{\sqrt{2}}\left(|01\rangle - |10\rangle\right).
\end{equation}
The crucial observation is that this entangled two-qubit state can be transformed into 
four orthogonal states of the two-qubit Hilbert space by performing unitary operations on just a single qubit. 
For instance, Alice can apply a rotation (the Pauli operations)  or do nothing to her part of the singlet, while Bob does nothing, to obtain the three triplets (or the singlet):
\begin{eqnarray}
\sigma_x \otimes I |\Psi^-\rangle = - |\Phi^-\rangle, & \quad & \sigma_y \otimes  I |\Psi^-\rangle = i |\Phi^+\rangle, \nonumber\\
\sigma_z \otimes I |\Psi^-\rangle =  |\Psi^+\rangle,  & \quad  & I \otimes I |\Psi^-\rangle      =  |\Psi^-\rangle, 
\end{eqnarray}
where
\begin{eqnarray}
\label{gorur-garir-headlight}
|\Psi^\pm\rangle &=& \frac{1}{\sqrt{2}}\left(|01\rangle \pm |10\rangle\right), \nonumber\\
|\Phi^\pm\rangle &=& \frac{1}{\sqrt{2}}\left(|00\rangle \pm |11\rangle\right), 
\end{eqnarray}
 are the Bell states and \(I\) is the qubit identity operator.  Suppose that the classical information that Alice wants to send to Bob is \(i\), where 
\(i= 0,1,2,3\). Alice and Bob previously agree on the following correspondence between the operations applied at Alice's end and 
the information \(i\) that she wants to send:
\begin{eqnarray}
\sigma_x \Rightarrow i=0, & \quad & \sigma_y \Rightarrow i=1, \nonumber\\
\sigma_z \Rightarrow i=2, & \quad & I \Rightarrow i=3.
\end{eqnarray} 
Depending on the classical information she wishes to send, Alice applies the appropriate rotation on her part of the shared singlet, according 
to the above correspondence. Afterwards, Alice sends her part of the shared state to Bob, via the noiseless 
quantum channel. Bob now has in his possession, the entire two-qubit state, which is in any of the four Bell states 
\(\left\{|\Psi^\pm\rangle, |\Phi^\pm\rangle \right\}\). Since these states are mutually orthogonal, he will 
be able to distinguish between them and hence find out the classical information sent by Alice.

To consider a more realistic scenario, usually two avenues are taken. One approach 
is to consider a \emph{noisy} quantum channel, while the additional resource is an arbitrary amount of shared bipartite \emph{pure}
state entanglement (see e.g. \cite{Bennett-ek, Bennett-dui}, see also \cite{MarieCurie, Debu}). 
The other approach
is to consider a \emph{noiseless} quantum channel, while the assistance is by a 
given bipartite \emph{mixed} entangled state (see e.g. \cite{MarieCurie,Debu,ek1,dui1, char1, tin1}).

Here, we consider the second approach, 
and derive 
the capacity of dense coding 
in this scenario, for a given state, where the the capacity is defined as the number of classical bits that can be
accessed 
by the receiver, per usage of the noiseless channel.
This will lead to a 
classification of bipartite states according to their ability to 
assist in dense coding. 
In the case where a noisy channel and an arbitrary amount of shared pure entanglement is considered, 
the capacity refers to the channel (see e.g. \cite{Bennett-ek, Bennett-dui}). However, in our case when a noiseless channel and a 
given shared (possibly mixed) state is considered, the capacity refers to the state. Note that the mixed shared state in our case can be thought 
of as an output of a noisy channel. 
A crucial element in finding the capacity of dense coding 
is the Holevo bound \cite{ref-halum}, which 
is a universal upper bound on classical information that can be decoded from a quantum ensemble. Below we  
discuss the bound, 
and subsequently derive the capacity of dense coding.

\subsection{The Holevo bound} \index{Holevo bound}
\label{sec-halum}

The Holevo bound is an upper bound on the amount of classical information that can be accessed from a quantum ensemble in which the information is 
encoded. 
Suppose therefore that Alice (\(A\)) obtains the classical message \(i\) that occurs with probability \(p_i\), 
and she wants to send it to Bob (\(B\)). Alice encodes this information \(i\) in a quantum state \(\rho_i\), and sends it to Bob. 
Bob receives the ensemble \(\{p_i, \rho_i\}\), and wants to obtain as much information as possible about \(i\). To do so, 
he performs a measurement, that gives the result \(m\), with probability \(q_m\). Let the corresponding post-measurement ensemble be 
\(\{p_{i|m}, \rho_{i|m}\}\). The information gathered can be quantified by the mutual information between the 
message index \(i\) and the measurement outcome \cite{Chennai}:
\begin{equation}
I(i:m)= H(\{p_i\}) - \sum_m q_m H(\{p_{i|m}\}).
\end{equation}
Note that the mutual information can be seen as the difference between the initial disorder and the (average) final disorder. 
Bob will be interested to obtain the maximal information, which is maximum of \(I(i:m)\) for all measurement strategies. This 
quantity is called the accessible information:
\begin{equation}
I_{acc} = \max I(i:m),
\end{equation} 
where the maximization is over all measurement strategies.

The maximization involved in the definition of accessible information is usually hard to compute, and hence the importance of bounds 
\cite{ref-halum,Utpakhi}. In particular, in Ref. \cite{ref-halum}, a universal upper bound, the Holevo bound, on \(I_{acc}\) is 
given: 
\begin{equation}
I_{acc}(\{p_i, \rho_i\}) \leq \chi(\{p_i, \rho_i\}) \equiv S(\overline{\rho}) - \sum_i p_i S(\rho_i).
\end{equation}
See also \cite{Rajabazar1,Khajuraho,Rajabazar2}. Here \(\overline{\rho} = \sum_ip_i\rho_i\) is the average ensemble state, and 
\(S(\varsigma)= - \mbox{tr}(\varsigma \log_2 \varsigma)\) 
is the von Neumann entropy of \(\varsigma\).

The Holevo bound is asymptotically  achievable in the sense that if the sender Alice is able to wait long enough and send 
long strings of the input quantum states \(\rho_i\), then there exists a particular encoding and a decoding scheme that asymptotically attains 
the bound. Moreover, the encoding consists in collecting certain long and ``typical'' strings of the input states, and sending them all at once 
\cite{babarey, maarey}.

\subsection{Capacity of quantum dense coding} \index{Capacity of quantum dense coding}
\label{koto-capa-re?}

Suppose that Alice and Bob share a quantum state \(\rho_{AB}\). Alice performs the unitary operation \(U_i\) with probability \(p_i\), on her 
part of the state \(\rho_{AB}\). The classical information that she wants to send to Bob is \(i\). 
Subsequent to her unitary rotation, she sends her part of the state 
\(\rho^{AB}\) to Bob. Bob then has the ensemble \(\{p_i, \rho_i\}\), where 
\[\rho_i = U_i \otimes I \rho_{AB} U_i^\dagger \otimes I.\]

The information that Bob is able to gather is \(I_{acc}(\{p_i, \rho_i\})\). This quantity is bounded above by 
\(\chi(\{p_i, \rho_i\})\), and is asymptotically achievable. 
The ``one-capacity'' \(C^{(1)}\) of dense coding for the state \(\rho_{AB}\) is the Holevo bound for the best encoding by Alice:
\begin{equation}
\label{Kohinoor}
C^{(1)}(\rho) = \max_{p_i,U_i} \chi(\{p_i, \rho_i\}) \equiv \max_{p_i,U_i} \left(S(\overline{\rho}) - \sum_i p_i S(\rho_i)\right).
\end{equation}
The superscript \((1)\) reflects the fact that Alice is using the shared state once at a time, during the asymptotic process. 
She is not using entangled unitaries on more than one copy of her parts of the shared states \(\rho_{AB}\). As we will see below, 
encoding with entangled unitaries does not help her to send more information to Bob.

\begin{figure}
\begin{center}
\includegraphics[scale=0.3]{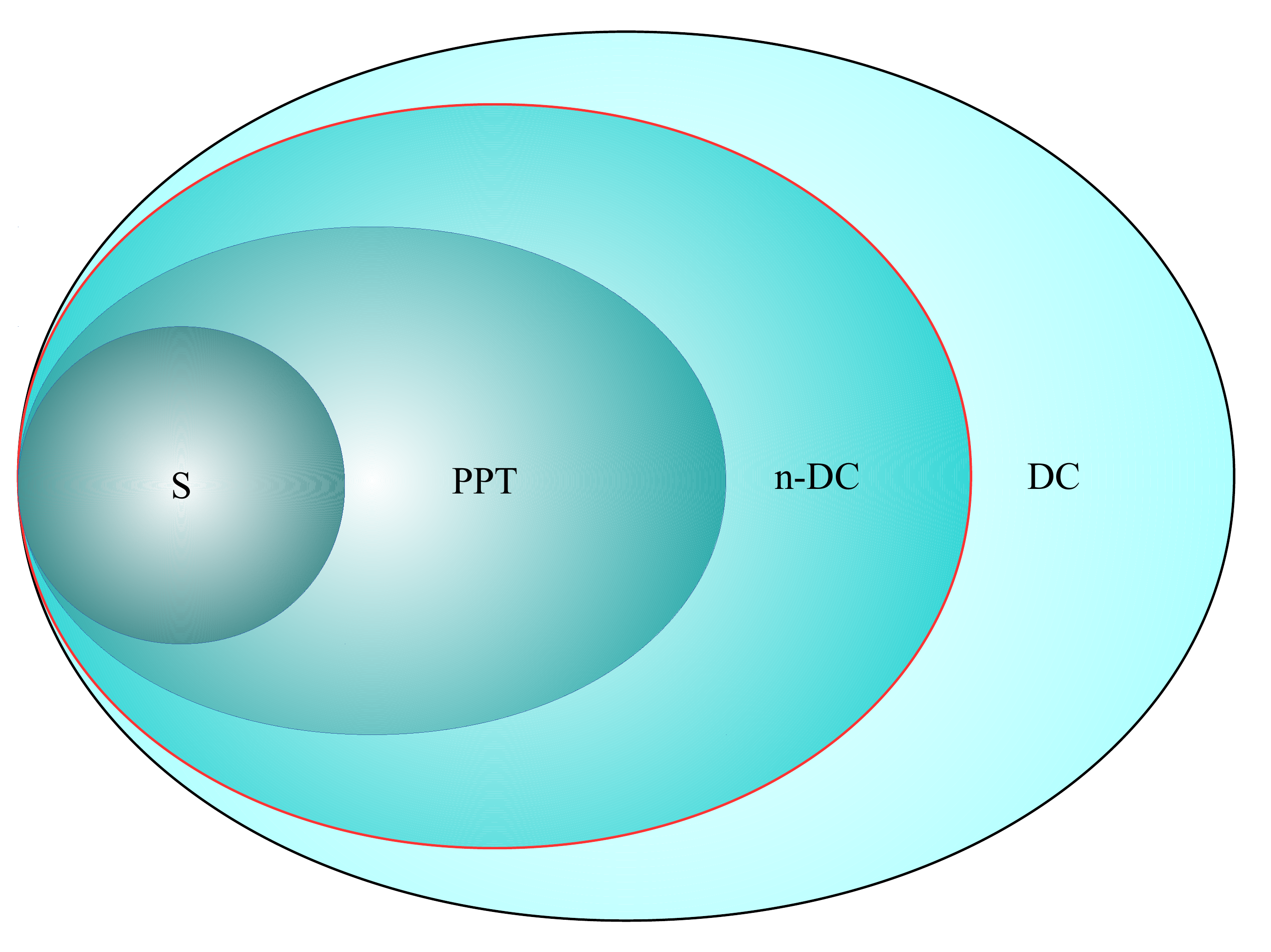}
\end{center}
\caption[Classification of bipartite quantum states according to their usefulness in dense coding]
{Classification of bipartite quantum states according to their usefulness in dense coding. The convex innermost 
region, marked as S, consists of separable states. The shell surrounding it, marked as PPT, is the set of 
PPT entangled states. The next shell, marked as n-DC, is the set of all states that are NPT, but not useful for dense 
coding. The outermost shell is that of dense-codeable states.}\label{fig:ghorar-dim}
\end{figure}

In performing the maximization in Eq. (\ref{Kohinoor}), first note that the second term in the right hand side (rhs) is 
\(-S(\rho)\), for all choices of the unitaries and probabilities. Secondly, we have 
\[S(\overline{\rho}) \leq S(\overline{\rho}_A) 
+ S(\overline{\rho}_B) \leq \log_2 d_A + S(\overline{\rho}_B),\] where 
\(d_A\) is the dimension of Alice's part of the Hilbert space of \(\rho_{AB}\), and 
 \( \overline{\rho}_A = \mbox{tr}_B \overline{\rho}\), \( \overline{\rho}_B = \mbox{tr}_A \overline{\rho}\). 
Moreover, \(S(\overline{\rho}_B) = S(\rho_B)\),
as nothing was done at Bob's end during the encoding procedure. (In any case, unitary operations does not change the spectrum, and hence 
the entropy, of a state.) Therefore, we have 
\[
\max_{p_i,U_i} S(\overline{\rho}) \leq \log_2 d_A + S(\rho_B).
\]
But the bound is reached by any 
complete set of orthogonal unitary operators
$\{W_j\}$, to be  chosen with equal probabilities, 
which 
satisfy the \emph{trace rule} $\frac{1}{d_A^2}\sum_j W_j^\dagger \Xi W_j=\mbox{tr}[\Xi]I$, 
for any operator
$\Xi$. Therefore, we have 
\[
C^{(1)}(\rho) = \log_2 d_A + S( \rho_B ) - S(\rho). 
\]

The optimization procedure above sketched essentially follows that in Ref. \cite{tin1}.
Several other lines of argument are possible for the maximization. One is given in Ref. \cite{dui1} (see also \cite{TajMahal}).  Another way to proceed is 
to guess where the maximum is reached (maybe from examples or by taking the most symmetric option), and then perturb the guessed result. If the 
first order perturbations vanish, the guessed result is correct, as the von Neumann entropy is a concave function and 
the maximization is carried out over a continuous parameter space. 

Without using the additional resource of entangled states, Alice will be able to reach a capacity of just 
\(\log_2d_A\) bits. Therefore, entanglement in a state \(\rho_{AB}\) is useful for dense coding if \(S( \rho_B ) - S(\rho)>0\). 
Such states will be called dense-codeable (DC) states. Such states exist, an example being the singlet state.

Note here that if Alice is able to use entangled unitaries on two copies of the shared state \(\rho\), 
the capacity is not enhanced (see Ref. \cite{ar-keu-korechhey?}). Therefore, the one-capacity is really the asymptotic
capacity, in this case. Note however that this additivity is known 
only in the case of encoding by unitary operations. A more general encoding may still have 
additivity problems (see e.g. \cite{Debu}). Here, 
we have considered unitary encoding only. This case is both mathematically more accessible, and experimentally more viable.


A bipartite state \(\rho_{AB}\) is useful for dense coding if and only if 
\(S( \rho_B ) - S(\rho)>0\). It can be shown that this relation cannot hold for PPT entangled states \cite{MarieCurie} (see also
\cite{TajMahal}).
Therefore a DC state is always NPT. 
However, the converse is not true: There exist states which are NPT, but not useful for dense coding. Examples of 
such states can be obtained by the considering the Werner state
\(\rho^W_{AB}(p) = p|\Psi^-\rangle \langle \Psi^- | + \frac{1-p}{4} I \otimes I\) \cite{Werner}.

The discussions above leads to the following classification of bipartite quantum states: 
\begin{enumerate}
\item Separable states: These states are of course not useful for dense coding. They can be prepared by LOCC.

\item PPT entangled states: These states, despite being entangled, cannot be used for dense  coding. 
Moreover,
their entanglement cannot be detected by the  partial transposition criterion.

\item NPT non-DC states: These states are entangled, and their entanglement can be detected by the 
partial transposition criterion. However, they are not useful for dense coding. 

\item DC states: These entangled states can be used for dense coding. 

\end{enumerate}
The above classification is illustrated in figure \ref{fig:ghorar-dim}.
A generalization of this classification has been considered in Refs. \cite{TajMahal,ar-keu-korechhey?}.

\section{Multipartite states} \index{Multipartite states}
\label{sec:multi}

The discussion about detection of bipartite entanglement presented above is of course quite far from 
complete. And yet, in this section, we present a few remarks on multipartite states and multipartite entanglement.
%

The case of detection of entanglement of pure states is again simple, although there are different types of entanglement present in a multipartite system. One quickly realizes that a multipartite pure state 
is entangled if and only if it is entangled in at least one bipartite splitting. 
So, for example, the  Greenberger-Horne-Zeilinger (GHZ) state \index{Greenberger-Horne-Zeilinger (GHZ) state} \(|\mbox{GHZ}\rangle = \frac{1}{\sqrt{2}} ( |000\rangle + |111\rangle ) \) \cite{GHZ}, shared between three parties $A$, $B$ and $C$, is 
entangled, because it is entangled in the $A$:$BC$ bipartition (as also in all others), whereas the state 
\(|\phi\rangle = \frac{1}{\sqrt{2}} (|000\rangle + |101\rangle) \) is entangled in the $A$:$BC$ and $AB$:$C$ bipartite splits but not in the $B$:$AC$ one. 

\begin{figure}
\includegraphics[scale=0.3]{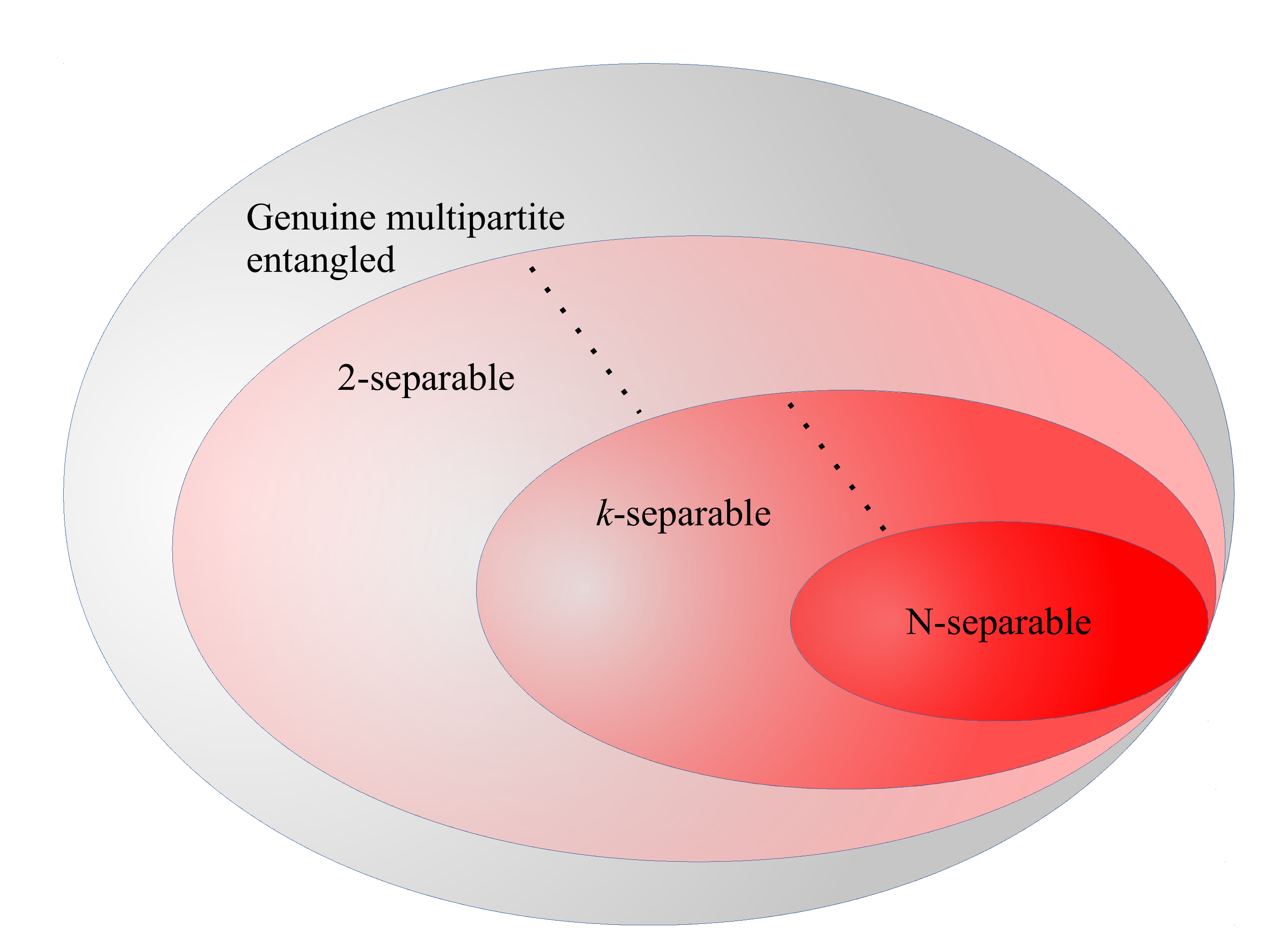}
\caption{ Geometric representation of the hierarchy of multipartite entangled states among  $N$-party  quantum
states. Each red-shaded set contains $k$-separable states (for $2 \leq k \leq N$), while the grey-shaded set contains all genuinely multipartite entangled states. The dotted lines indicate that there can be several $k$-separable state sets between $2$-separable and $N$-separable ones.}
\label{fig:p_sep}
\end{figure}

\subsection{$k$-separable, fully-separable, and genuine multipartite entangled states} \index{$k$-separability}
\label{sec:ksep}
Among multipartite states, there exists a hierarchical structure of states with respect to their entanglement quality. An $N$-party pure quantum state is called $k$-separable ($2 \leq k \leq N$), if it is separable in at least $k-1$ bipartite splitting. Similarly, an $N$-party pure quantum state is $N$-separable or fully-separable if it is separable in all bipartite splittings. A pure quantum state possesses genuine multipartite entanglement if and only if it is entangled in all possible bipartite cuts. For example, the state \(|\phi\rangle = \frac{1}{\sqrt{2}} (|000\rangle + |101\rangle) \) is bi-separable or $2$-separable, whereas  the GHZ state is genuinely multipartite entangled. 

The case of mixed states is more involved. A possibly mixed quantum state $\rho$ of $N$ parties is $k$-separable, if in every pure state decomposition of $\rho = \sum_i p_i \Ket{\psi^i}\Bra{\psi^i}$, there exist at least one $k$-separable pure state and no other state with separability lesser than $k$. Similarly,  a possibly mixed quantum state is genuinely multipartite entangled, if it has at least one genuine multipartite entangled pure state in every pure state decomposition of it. For example, in the three qubit case, the equal mixture of the W state \index{W state} $\Ket{W} = \frac{1}{\sqrt{3}}(\Ket{001} + \Ket{010} + \Ket{100})$ \cite{dur_vidal_cirac, W_state}  and its ``complement" $\Ket{\bar{W}} = \frac{1}{\sqrt{3}}(\Ket{110} + \Ket{101} + \Ket{011})$ is genuinely multipartite entangled \cite{qc_wt_cc}, and the equal mixture of $\Ket{\psi_1} = \frac{1}{\sqrt{2}}(\Ket{001} + \Ket{010})$ and $\Ket{\psi_2} = \frac{1}{\sqrt{2}}(\Ket{001} + \Ket{100})$ is bi-separable.
Figure \ref{fig:p_sep} depicts the schematic geometric picture of this hierarchical structure of multipartite entanglement.

One avenue to  quantify the degree of such multipartite entanglement relies on the above geometric structure of multipartite entangled states \cite{GM1, GM2, GM3}. Given a distance functional $\mathcal{D}$, that satisfies conditions (1)-(3) given in Sec. \ref{sec:rel_ent}, the quantity \index{Geometric measure of multipartite entanglement}
\begin{equation}
\mathcal{E}^{\mathcal{D}}_k(\rho) = \underset{\sigma \in k\mbox{-sep}}{\mbox{min}} \ \mathcal{D}(\rho, \sigma),
\label{eq:gm}
\end{equation}
gives a measure of $k$-\emph{inseparable} multipartite entanglement in the state $\rho$.
As two special cases, for $k=N$, Eq. (\ref{eq:gm}) gives the minimum distance from fully-separable states, and thus quantifies the ``total" multipartite entanglement, while for $k=2$, $\mathcal{E}^{\mathcal{D}}_2$ gives a  measure of genuine multipartite entanglement. Optimization in Eq. (\ref{eq:gm}) is a formidable problem for general multipartite states. But there exist forms of the geometric measure $\mathcal{E}^{\mathcal{D}}_k(\rho)$ for various  families of states (pure and mixed) corresponding to certain distance measures \cite{GGM1, GGM2, GGM3}. For example, in case of pure states, if we take the following distance measure
\begin{equation}
\mathcal{D}(\psi, \phi) = 1 - |\langle\psi|\phi\rangle|^2,
\end{equation}
we get the ``geometric measures" of multipartite entanglement for  pure states \cite{GM1, GM3, GGM1, GGM2}. In case the minimum distance is from bi-separable states, the corresponding measure has been termed as the generalized geometric measure \index{Generalized geometric measure} \cite{GGM1,GGM2},
\begin{equation}
\mathcal{G}(\psi) = \underset{\Ket{\phi} \ \in \ 2\mbox{-sep}}{\mbox{min}} \left(1-|\langle\psi|\phi\rangle|^2\right),
\label{eq:ggm1}
\end{equation}
which measures the genuine multiparty entanglement in $\Ket{\psi}$.
In this case, we get a computable form of the measure for an arbitrary  $N$-party pure state, $\Ket{\psi}$, shared between $A_1, A_2, ..., A_N$, in arbitrary dimensions, given by \cite{GGM2}
\begin{equation}
\mathcal{G}(\psi) = 1 - \mbox{max}\{\lambda^2_{A:B} | A  \cup B = A_1, A_2, ..., A_N, A \cap B = \phi\},
\label{eq:ggm2}
\end{equation}
where $\lambda_{A:B}$ is the maximum Schmidt coefficient in each possible bipartition split of the type $A : B$ of the given state $\Ket{\psi}$.

Until now, in the case of multiparty mixed states, we have considered only the distance-based measures. However, it is also possible to use the convex-roof approach to define entanglement measures for multiparty mixed states, after choosing a certain measure for pure states \cite{GGM4, GGM5}.


\subsection{Three qubit case: GHZ class vs. W class} 
\begin{figure}
\includegraphics[scale=0.3]{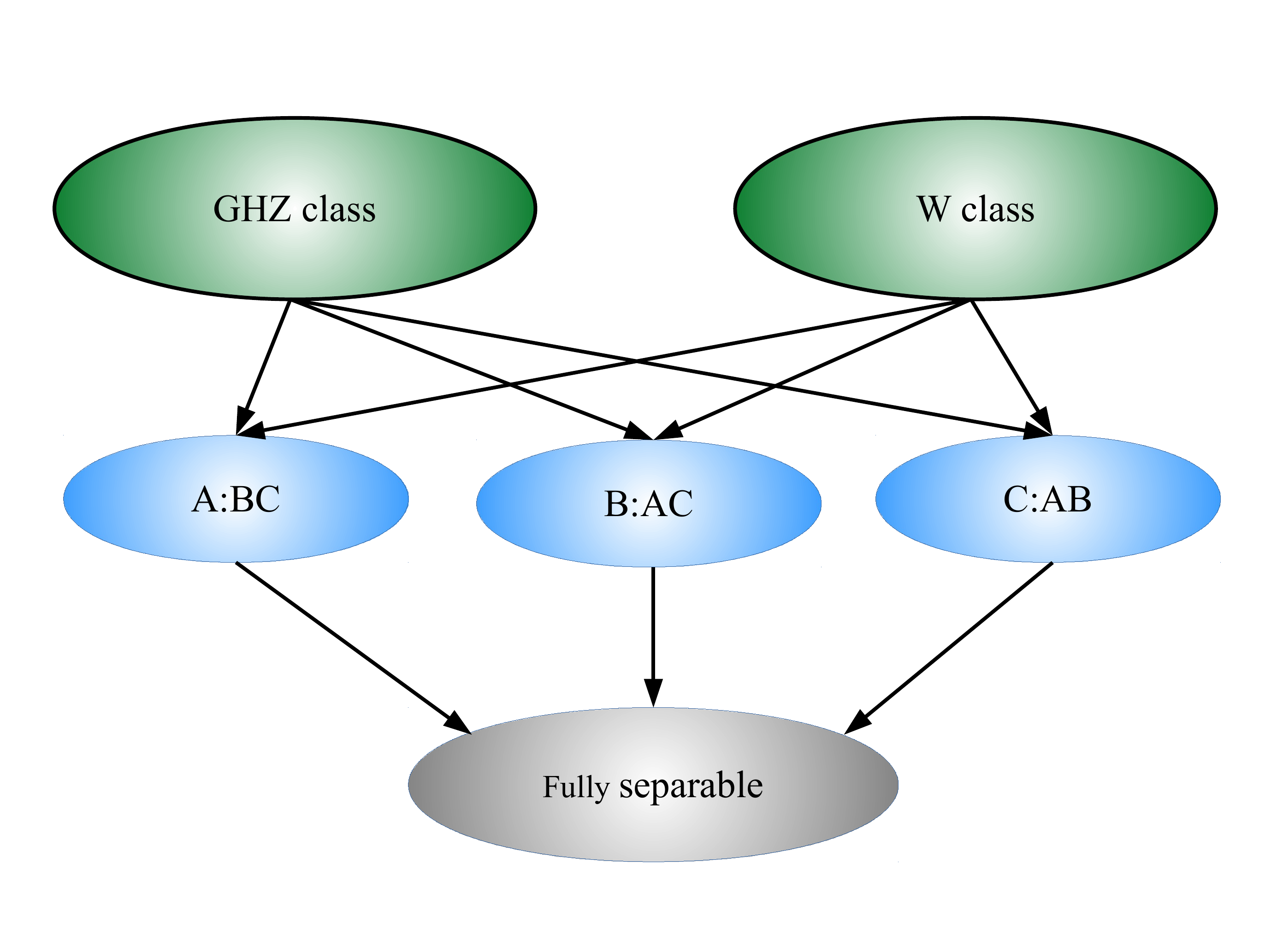}
\caption{ Different classes of three-qubit pure states. Two states in the same class are SLOCC equivalent, i.e., those can be converted to one another under SLOCC operations. The direction of the arrows shows which non-invertible transformations
between classes are possible via SLOCC.  Reproduced figure with permission from the Authors of Ref. \cite{dur_vidal_cirac}. Copyright (2000) of the American Physical Society.}
\label{fig:three_qubit}
\end{figure}

\index{GHZ class} \index{W class}
For three-qubit pure states, the above classification of multipartite states boils down to three categories, namely 
\begin{enumerate}
\item fully-separable states of the form $\Ket{\psi_A} \otimes \Ket{\psi_B} \otimes \Ket{\psi_C}$, 
\item bi-separable states of three types: $A$:$BC$, $B$:$AC$, and $C$:$AB$, where $A$:$BC$ type states are separable in the $A$:$BC$ splitting but not in others and so on,
\item and finally, genuine tripartite entangled states.
\end{enumerate}
Another classification is possible by considering interconversion of  states through stochastic local
operations and classical communication (SLOCC) \index{stochastic local
operations and classical communication (SLOCC)} \cite{slocc}, i.e, through LOCC but with a non-unit probability. In this scenario, we call two states $\Ket{\psi}$ and $\Ket{\phi}$ to be equivalent if there is a non-vanishing
probability of success when trying to convert $\Ket{\psi}$ into $\Ket{\phi}$ as well as in the opposite direction through SLOCC. For example, in the two-qubit case, every entangled state is equivalent to any other entangled states, and the entropies of entanglement quantify these conversion rates (see Sec. \ref{sec:ent_measures}). This happy situation is absent already in the case of pure three-qubit states \cite{dur_vidal_cirac}. It turns out that any genuine three-qubit pure entangled state can be converted into either the GHZ state  or the W state, but not both, using SLOCC. This divides the set of genuine three-qubit pure entangled
states into two sets which are incompatible under SLOCC. In other words, if a state $\Ket{\psi}$ is convertible into $\Ket{GHZ}$ and another state $\Ket{\phi}$ is convertible into $\Ket{W}$ via SLOCC, then one cannot transform $\Ket{\psi}$ to $\Ket{\phi}$ or vice-versa, with any non-zero probability. These two sets of genuine three-qubit pure entangled states are termed as the GHZ-class and the W-class respectively. Figure \ref{fig:three_qubit} shows these different classes of three-qubit pure states and possible SLOCC transformations between the different classes.

D\"ur  et al. \cite{dur_vidal_cirac} presented general forms of the members of each class. A member of the GHZ-class can be expressed as
\begin{equation}
\Ket{\psi_{GHZ}} = \sqrt{K}(c_\delta|0\rangle|0\rangle|0\rangle+s_\delta e^{i\varphi}|\varphi_A\rangle|\varphi_B\rangle|\varphi_C\rangle),
\end{equation}
where
\begin{eqnarray}
\Ket{\varphi_A}&=&c_{\alpha}\Ket{0}+s_{\alpha}\Ket{1},\nonumber\\
\Ket{\varphi_B}&=&c_{\beta}\Ket{0}+s_{\beta}\Ket{1},\nonumber\\
\Ket{\varphi_C}&=&c_{\gamma}\Ket{0}+s_{\gamma}\Ket{1},
\end{eqnarray}
 $K=(1+2 c_\delta s_\delta c_\alpha c_\beta c_\gamma c_\varphi)^{-1} \in
(1/2,\infty)$ is a normalization factor, and $\delta \in (0,\pi/4]$, $\alpha,\beta,\gamma \in (0,\pi/2]$ and $\varphi \in
[0,2\pi)$. Here $c_{\alpha} = \cos \alpha$, $s_{\alpha} = \sin \alpha$, etc. Similarly, a member state of the W-class, upto a local 
unitary transformation, can be written as,
\begin{equation}
\Ket{\psi_W} = \sqrt{a} \Ket{001} +\sqrt{b} \Ket{010} + \sqrt{c} \Ket{100} +\sqrt{d} \Ket{000},
\end{equation}
where $a, b, c >0$, and $d = 1- (a + b+ c) \geq 0$.

For multipartite states with $N \geq 4$, there exist infinitely many inequivalent kinds of such entanglement classes under SLOCC \cite{dur_vidal_cirac}. See Refs. \cite{frank_9, miyake} for further results.

\subsection{Monogamy of quantum entanglement} \index{Monogamy of quantum entanglement}
The concept of monogamy \cite{bennett_eof1, terhal_mono, ckw, mono_review} is an inherent feature of multipartite quantum correlations, and in particular, of sharing of two-party entanglements in multiparty quantum states. Unlike classical correlations, quantum entanglement cannot be freely shared among many parties.  For example, given three  parties $A$, $B$, and $C$, if party $A$ is maximally entangled with party $B$, e.g., if they share a singlet state $\Ket{\Psi^-} = (\Ket{01} - \Ket{10})/\sqrt{2}$, then $A$ cannot be simultaneously entangled with party $C$. In other words, there exist a trade-off between $A$'s entanglement with $B$ and its entanglement with $C$. 
In principle, and in its simplest form, for a two-party entanglement measure $\mathcal{E}$ and a three party system shared between $A$, $B$, and $C$, any relation providing an upper bound to the sum $\mathcal{E}_{A:B} + \mathcal{E}_{A:C}$ that is stronger than the sum of individual maxima of $\mathcal{E}_{A:B}$ and $\mathcal{E}_{A:C}$, can be termed as a monogamy relation for $\mathcal{E}$. However, in Ref. \cite{ckw}, an intuitive reasoning for the validity of the relation 
\begin{equation}
\mathcal{E}_{A:B} + \mathcal{E}_{A:C} \leq \mathcal{E}_{A:BC},
\label{eq:monogamy}
\end{equation}
is given. As in Ref. \cite{ckw}, we will call an entanglement measure $\mathcal{E}$ to be monogamous in a certain three-party system, if the relation (\ref{eq:monogamy}) is valid for all quantum states in that system.

\noindent \textbf{Theorem 17} \cite{ckw} \\
\emph{For an arbitrary three-qubit pure state shared between $A$, $B$, and $C$, the squared concurrence $\mathcal{C}_{A:B}^2$ between $A$ and $B$, plus the squared concurrence $\mathcal{C}_{A:C}^2$ between $A$ and $C$, cannot be greater than the squared concurrence $\mathcal{C}_{A:BC}^2$ between $A$ and the pair $BC$.}

It was shown later that $(\mathcal{C}^2_{A:B} + \mathcal{C}^2_{A:C}) \leq \mathcal{C}^2_{A:BC}$ holds also for arbitrary mixed three-qubit 
states \cite{frank_mono}, where $\mathcal{C}^2_{A:BC}$ is defined via convex-roof extension.
Using Theorem 17, one can define a positive quantity in terms of squared concurrence, named \emph{tangle}, or \emph{three-tangle}, for three-qubit pure states, as \cite{ckw} \index{Tangle} 
\begin{equation}
\tau_{ABC} = \mathcal{C}^2_{A:BC} - (\mathcal{C}^2_{A:B} + \mathcal{C}^2_{A:C}).
\label{eq:tangle}
\end{equation}
The tangle $\tau_{ABC}$, also called ``residual entanglement", is independent of the choice of the ``node" or ``focus", which is the party A here. It has been argued that the tangle $\tau_{ABC}$ gives a quantification of three-qubit entanglement. The generalization of the tangle to mixed states can e.g. be obtained by the convex roof extension, which is difficult to compute.
The tangle is a proper entanglement monotone, as it does not increase on average under LOCC \cite{dur_vidal_cirac}. It also successfully distinguishes the two inequivalent SLOCC classes in three-qubit pure state scenario, namely the GHZ-class and the W-class. It has been shown that tangle vanishes for states in the W-class, whereas it is always non-zero for  states in the GHZ-class \cite{dur_vidal_cirac}.
Therefore, to quantify entanglement content of states from the W-class, one has to look for other multipartite entanglement measures, different from tangle.

In the $N$-party scenario, generalization of inequality (\ref{eq:monogamy}) can be written as
\begin{equation}
\mathcal{E}_{A_1:A_2} + \mathcal{E}_{A_1:A_3} + ... + \mathcal{E}_{A_1:A_N} \leq \mathcal{E}_{A_1:A_2A_3...A_N}.
\label{eq:monogamy2}
\end{equation}  
In the same spirit as for the definition of the tangle in Eq. (\ref{eq:tangle}), we can define the ``monogamy score" \cite{mono_score} corresponding to a bipartite entanglement measure $\mathcal{E}$ as \index{Monogamy score}
\begin{equation}
 \delta^{A_1}_{\mathcal{E}} = \mathcal{E}_{A_1:A_2A_3...A_N} - (\mathcal{E}_{A_1:A_2} + \mathcal{E}_{A_1:A_3} + ... + \mathcal{E}_{A_1:A_N}),
\label{eq:mono_score}
\end{equation}  
with party $A_1$ as ``nodal". Like for the tangle, it has been argued that the monogamy score, $\delta^{A_1}_{\mathcal{E}}$, can act as a measure of multiparty entanglement \cite{ckw, mono_score}, obtained by subtracting the bipartite contributions $(\mathcal{E}_{A_1:A_2} , \mathcal{E}_{A_1:A_3} , ... , \mathcal{E}_{A_1:A_N})$ in the ``total" entanglement $\mathcal{E}_{A_1:A_2A_3...A_N}$ in the $A_1:A_2A_3...A_N$ partition. Unlike the tangle, monogamy scores for certain entanglement measures can possess negative values for some $N$-party quantum states.
For further information on recent works about the monogamy of quantum entanglement and correlations, see Refs. \cite{mono_review} and references therein. \\

For further 
results about entanglement criteria, detection, and classification of multipartite states,  
see e.g. 
\cite{panch, saat, Mumbai, satattor, Horo-realign, sakkhiprl, ek, chhoi, char, tin, dui, sotero}, and 
references therein.

\section{Problems}

\noindent \textbf{Problem 1} Show that the singlet state $\Ket{\Psi^-} = \frac{1}{\sqrt{2}} (\Ket{01} - \Ket{10})$ has non-positive partial transposition.

\noindent \textbf{Problem 2} Consider the Werner state  \(\rho^W_{AB}(p) = p |\Psi^- \rangle \langle \Psi^- | + (1 -p) I/4\) in \(\mathbb{C}^2 \otimes \mathbb{C}^2\), where 
\(0\leq p \leq 1\) \cite{Werner}. Find the values of the mixing parameter \(p\),
for which entanglement in the Werner state can be detected by the partial transposition criterion.

\noindent \textbf{Problem 3} Show that in \({\mathbb{C}}^2 \otimes {\mathbb{C}}^2\), the partial transposition of a 
density matrix can have at most one negative eigenvalue. 

\noindent \textbf{Problem 4} Given two random variables \(X\) and \(Y\), show that the Shannon entropy of the joint distribution cannot 
be smaller than that of either.

\noindent \textbf{Problem 5} Prove Theorem 5. 

\noindent \textbf{Problem 6} Consider the following state in $\mathbb{C}^{3}\otimes\mathbb{C}^{3}$:
\begin{equation}
 \rho_{AB}(\alpha) = \frac{2}{7}|\psi\rangle\langle \psi| + \frac{\alpha}{7}\varrho_+ + \frac{5 -\alpha}{7}\varrho_-,
 \label{3x3state2}
\end{equation}
where $ \varrho_+=(|01\rangle\langle 01|+|12\rangle\langle 12|+|20\rangle\langle 20|)/3$, 
$\varrho_-=(|10\rangle\langle 10| + |21\rangle\langle 21| + |02\rangle\langle 02|)/3$, 
$|\psi\rangle = \frac{1}{\sqrt{3}}\sum_{i = 0}^2 |ii\rangle$, and $0\leq\alpha\leq5$  \cite{be_horo}.
Find the ranges of the  parameter \(\alpha\), for which entanglement in the state $\rho_{AB}(\alpha)$ can be detected by the majorization and the cross-norm  criteria.

\noindent \textbf{Problem 7} Prove Corollaries 1 and 2.

\noindent \textbf{Problem 8} Prove Lemma 1. 

\noindent \textbf{Problem 9} Prove Theorem 10.

\noindent \textbf{Problem 10} Consider the the following set of orthogonal product states in $\mathbb{C}^3 \otimes \mathbb{C}^3$ \cite{UPBPRL}:
\begin{eqnarray}
{|\psi _{0}\rangle } &=&{\frac{1}{\sqrt{2}}}{|0\rangle } \otimes ({|0\rangle }-{\
|1\rangle }),\ \ {|\psi _{1}\rangle }={\frac{1}{\sqrt{2}}}({|0\rangle }-{\
|1\rangle })\otimes{|2\rangle }, \nonumber \\ 
{|\psi _{2}\rangle } &=&{\frac{1}{\sqrt{2}}|2\rangle }\otimes({|1\rangle }-{\
|2\rangle }),\text{ \ }{|\psi _{3}\rangle }={\frac{1}{\sqrt{2}}} \otimes ({|1\rangle }
-{|2\rangle }){|0\rangle }, \nonumber \\
{|\psi _{4}\rangle } &=&{\frac{1}{3}}({|0\rangle }+{|1\rangle }+{|2\rangle )} \otimes
({|0\rangle }+{|1\rangle }+{|2\rangle )}.
\end{eqnarray}
Using the range criterion, show that the quantum state 
\begin{equation}
\rho =\frac{1}{4}(I-\sum_{i=0}^{4}{|\psi _{i}\rangle \langle \psi
_{i}|})
\end{equation}
is entangled, where $I$ denotes the identity operator on $\mathbb{C}^3 \otimes \mathbb{C}^3$.

\noindent \textbf{Problem 11} Prove Theorem 15.

\noindent \textbf{Problem 12} Find the relative entropy of entanglement for Werner state \(\rho^W_{AB}(p) = p |\Psi^- \rangle \langle \Psi^- | + (1 -p) I/4\) in \(\mathbb{C}^2 \otimes \mathbb{C}^2\), where 
\(0\leq p \leq 1\) \cite{Werner}.

\noindent \textbf{Problem 13} Show that each of the shells depicted in figure \ref{fig:ghorar-dim} are nonempty, and of 
nonzero measure. Show also that all the boundaries are convex. 

\noindent \textbf{Problem 14} Show that the entanglement of formation is non-monogamous for three-qubit states.

\noindent \textbf{Problem 15} Consider the geometric measure of genuine multipartite entanglement $\mathcal{G}(\psi)$ given in Eq. (\ref{eq:ggm1}), and then show that it assumes the computable closed form given in Eq. (\ref{eq:ggm2}).

\section*{Acknowledgements}

ML  acknowledge financial support from the John Templeton Foundation, the EU grants OSYRIS (ERC-2013-AdG Grant No. 339106), QUIC (H2020-FETPROACT-2014 No. 641122), and SIQS (FP7-ICT-2011-9 No. 600645), the Spanish MINECO grants FOQUS (FIS2013-46768-P),  F\"ISICATEAMO (FIS2016-79508-P), and ``Severo Ochoa" Programme (SEV-2015-0522), the Generalitat de Catalunya  support (2014 SGR 874) and CERCA/Programme, and Fundació Privada Cellex. AS acknowledges financial support from the Spanish MINECO projects FIS2013-40627-P,FIS2016-80681-P
the Generalitat de Catalunya CIRIT (2014-SGR-966).


\printindex

\begin{thebibliography}{999}

\bibitem{Ekert} A.K. Ekert, Phys. Rev. Lett. \textbf{67}, 661 (1991).

\bibitem{BW}C.H. Bennett and S.J. Wiesner, Phys. Rev. Lett. \textbf{69}, 2881 (1992).

\bibitem{BBCJPW}C.H. Bennett, G. Brassard, C. Crepeau, R. Josza, A Peres, and  W.K. Wootters, Phys. Rev. Lett. \textbf{70}, 1895 (1993).



\bibitem{ZZHE}M. \.Zukowski, A. Zeilinger, M.A. Horne and A.K. Ekert, Phys. Rev. Lett.
\textbf{71}, 4287 (1993); 
M. \.Zukowski, A. Zeilinger, and H. Weinfurter,
Annals N.Y. Acad. Sci. \textbf{755}, 91 (1995);
S. Bose, V. Vedral, and P.L. Knight, Phys. Rev. A \textbf{57}, 822 (1998);
S. Bose, V. Vedral, and P.L. Knight, Phys. Rev. A \textbf{60}, 194 (1999).


\bibitem{Werner}R.F. Werner, Phys. Rev. A \textbf{40}, 4277(1989).



\bibitem{Pawelbound}P. Horodecki, Phys. Lett. A \textbf{232}, 333 (1997).

\bibitem{karnas00} S. Karnas and M. Lewenstein, J. Phys. A \textbf{34}, 6919 (2001).


\bibitem{PeresPPT} A. Peres, Phys. Rev. Lett \textbf{77}, 1413 (1996).
 

\bibitem{HorodeckiPPT} M. Horodecki, P. Horodecki, and R. Horodecki, 
Phys. Lett. A \textbf{223}, 1 (1996).


\bibitem{NielsenKempe} M.A. Nielsen and J. Kempe, Phys. Rev. Lett. \textbf{86}, 5184 (2001).

\bibitem{CCN1} O. Rudolph, J. Phys. A: Math. Gen. {\bf 33} 3951 (2000).

\bibitem{CCN2} K. Chen and L.-A. Wu, Quant. Inf. Comput., \textbf{3},  193 (2003).

\bibitem{CCN3} O. Rudolph, Quant. Info. Process., {\bf 4}, 219 (2005).

\bibitem{cmc1} O. G\"uhne, P. Hyllus, O. Gittsovich, and J. Eisert, Phys. Rev. Lett. {\bf 99}, 130504 (2007).

\bibitem{cmc2} O. Gittsovich, O. G\"uhne, P. Hyllus, and J. Eisert, Phys. Rev. A {\bf 78}, 052319 (2008).

\bibitem{reduction} M. Horodecki and P. Horodecki, Phys. Rev. A \textbf{59}, 4206 (2000).

\bibitem{Doherty} A. Doherty, P. Parillo, and F. Spedalieri, Phys. Rev. Lett. \textbf{88}, 187904 (2002);
Phys. Rev. A \textbf{69}, 022308 (2004);  F. Hulpke and D. Bru{\ss},
J. Phys. A: Math. Gen. \textbf{38}, 5573 (2005).

\bibitem{other_criteria3} H. F. Hofmann and S. Takeuchi, Phys. Rev. A {\bf 68}, 032103 (2003).

\bibitem{other_criteria1} J. I. de Vicente, Quantum Inf. Comput. {\bf 7}, 624 (2007).

\bibitem{other_criteria2} J. I. de Vicente, J. Phys. A {\bf 41}, 065309 (2008).

\bibitem{other_criteria4} C.-J. Zhang, Y.-S. Zhang, S. Zhang, and G.-C. Guo, Phys. Rev. A {\bf 76}, 012334 (2007);
C.-J. Zhang, Y.-S. Zhang, S. Zhang, and G.-C. Guo, Phys. Rev. A {\bf 77}, 060301(R) (2008).



\bibitem{Horodeckibound} M. Horodecki, P. Horodecki, and  R. Horodecki, Phys. Rev. Lett.
\textbf{80}, 5239 (1998), and references therein.


\bibitem{koto-volume-re} K. \.Zyczkowski,
P. Horodecki, A. Sanpera, and M. Lewenstein,
Phys. Rev. A, \textbf{58}, 883 (1998); 
K. \.Zyczkowski, Phys. Rev. A \textbf{60}, 3496 (1999);
S. Szarek, Phys. Rev. A {\bf 72}, 032304 (2005), 
and references therein.

\bibitem{biyog}  D.P. DiVincenzo, P.W. Shor, J.A. Smolin, B.M. Terhal, and A.V. Thapliyal,
Phys. Rev. A \textbf{61}, 062312 (2000); W. D\"ur, J.I. Cirac, M. Lewenstein, and D. Bru{\ss},
\emph{ibid.} \textbf{61}, 062313 (2000).
\bibitem{bennett_dist} C. H. Bennett, G. Brassard, S. Popescu, B. Schumacher, J. A. Smolin, and W. K. Wootters, Phys. Rev. Lett. {\bf 76}, 722 (1996).
\bibitem{bennett_eof2} C. H. Bennett, D. P. DiVincenzo, J. A. Smolin, and W. K. Wootters, Phys. Rev. A {\bf 54}, 3824 (1996).


\bibitem{majorizationBhatia} A.W. Marshall and I. Olkin, \emph{Inequalities: Theory of Majorization
and Its Applications} (Academic Press, New York, 1979); P.M. Alberti and A. Uhlmann, \emph{Stochasticity
and Partial order: Doubly Stochastic Maps and Unitary Mixing} (Dordrecht, Boston, 1982); 
R. Bhatia, \emph{Matrix Analysis} (Springer, New York, 1997).




\bibitem{Hiroshima} T. Hiroshima, Phys. Rev. Lett. \textbf{91}, 057902 (2003).

\bibitem{horo_ent_sep} R. Horodecki and P. Horodecki, Phys. Lett. A {\bf 194}, 147 (1994).

\bibitem{local_filter} N. Gisin, Phys. Lett. A {\bf 210}, 151 (1996);
M. Horodecki, P. Horodecki, and R. Horodecki, Phys. Rev. Lett. {\bf 78}, 574 (1997);
A. Kent, N. Linden, and S. Massar, Phys. Rev. Lett. {\bf 83}, 2656 (1999); 
F. Verstraete, J. Dehaene, and B. De Moor, Phys. Rev. A {\bf 65}, 032308 (2002).



\bibitem{sakkhi12} O. G{\"u}hne, P. Hyllus, D. Bru{\ss}, A. Ekert, M. Lewenstein, C. Macchiavello, and A. Sanpera,
Phys. Rev. A \textbf{66}, 062305 (2002);  O. G{\"u}hne, P. Hyllus, D. Bru{\ss}, A. Ekert, M. Lewenstein, C. Macchiavello, and A. Sanpera,
J. Mod. Opt. \textbf{50}, 1079 (2003).
 
 
 
\bibitem{sakkhiprl}  M. Bourennane, M. Eibl, C. Kurtsiefer, S. Gaertner, H. Weinfurter, O G{\"u}hne, P. Hyllus, D. Bru{\ss}, 
M. Lewenstein, and A. Sanpera,
Phys. Rev. Lett. \textbf{92}, 087902 (2004).




\bibitem{alt:1985} H.W. Alt, \emph{Lineare Funktionalanalysis}, (Springer-Verlag, 1985).


\bibitem{Woronowicz} S.L. Woronowicz, Commun. Math. Phys. \textbf{51}, 243 (1976); 
P. Kryszynski and S.L. Woronowicz, Lett. Math. Phys. \textbf{3}, 319 (1979); M.D. Choi, Proc. Sympos. Pure Math. \textbf{38}, 583 (1982).


\bibitem{Terhal} B.M. Terhal, Lin. Alg. Appl. \textbf{323}, 61 (2001).

\bibitem{sakkhi-ager} M. Lewenstein, B. Kraus, J.I. Cirac, and P. Horodecki, Phys. Rev. A \textbf{62}, 052310 (2000);
 D. Bru{\ss}, J.I. Cirac, P. Horodecki, F. Hulpke, B. Kraus, M. Lewenstein, and A. Sanpera, J. Mod. Opt. \textbf{49}, 1399 (2002). 


\bibitem{jamiolkowski} E. C. G. Sudarshan, P. M. Mathews, and J. Rau, Phys. Rev. {\bf 121}, 920 (1961);
A. Jamio{\l}kowski, Rep. Math. Phys., \textbf{3}, 275 (1972); M.-D. Choi, Linear Alg. Appl. {\bf 10}, 285 (1975);
K. Kraus, \emph{States, Effects, and Operations: Fundamental Notions of Quantum Theory}, \emph{Lecture notes in Physics}, vol. 190 (Spring-Verlag, New York, 1983).

\bibitem{EPRparadox} A. Einstein, B. Podolsky, and N. Rosen, Phys. Rev. \textbf{57}, 777 (1935). 


\bibitem{Bell}J.S. Bell, Physics \textbf{1}, 195 (1964).



\bibitem{gerakol} E. Santos, 
Phys. Rev. Lett. \textbf{66}, 1388 (1991);  E. Santos, 
Phys. Rev. A \textbf{46}, 3646 (1992);
P.M. Pearle, Phys. Rev. D 
\textbf{2}, 1418 (1970);
J.F. Clauser and M.A. Horne, Phys. Rev. D 
\textbf{10}, 526 (1974);
P.G. Kwiat, P.H. Eberhard, A.M. Steinberg, and R.Y. Chiao, 
Phys. Rev. A \textbf{49}, 3209 (1994);
N. Gisin and B. Gisin, 
Phys. Lett. A \textbf{260}, 323 (1999);
S. Massar, S. Pironio, J. Roland, and B. Gisin, 
Phys. Rev. A \textbf{66}, 052112 (2002);
    R. Garcia-Patr\'on, J. Fiur\'asek, N.J. Cerf, J. Wenger, R. Tualle-Brouri, and Ph. Grangier,  
Phys. Rev. Lett. \textbf{93}, 130409 (2004);
N. Brunner, N. Gisin, V. Scarani, and C. Simon, Phys. Rev. Lett. {\bf 98}, 220403 (2007);
R. Garcia-Patr\'on, J. Fiur\'asek, and N. J. Cerf, Phys. Rev. A {\bf 71}, 022105 (2005);
T. V\'ertesi, S. Pironio, and N. Brunner, Phys. Rev. Lett. {\bf 104}, 060401 (2010);
N. Sangouard, J.-D. Bancal, N. Gisin, W. Rosenfeld, P. Sekatski, M. Weber, and H. Weinfurter, Phys. Rev. A {\bf 84}, 052122 (2011);
A. Cabello and F. Sciarrino, Phys. Rev. X {\bf 2}, 021010 (2012);
J. Larsson, J. Phys. A {\bf 47}, 424003 (2014).




\bibitem{Paris-ghyama} S.J. Freedman and J.S. Clauser, Phys. Rev. Lett. \textbf{28}, 938 (1972); 
A. Aspect, P. Grangier, and G. Roger, Phys. Rev. Lett. \textbf{47}, 460 (1981); 
\emph{ibid.} \textbf{49}, 91 (1982);
 A. Aspect, J. Dalibard, and G. Roger, \emph{ibid.} \textbf{49}, 1804 (1982); 
P.G. Kwiat, K. Mattle, H. Weinfurter, A. Zeilinger, A.V. Sergienko, and Y. Shih, Phys. Rev. Lett. \textbf{75}, 4337 (1995);
 G. Weihs, T. Jennewein, C. Simon, H. Weinfurter, and A. Zeilinger, Phys. Rev. Lett. \textbf{81}, 5039 (1998);
 W. Tittel, J. Brendel, B. Gisin, T. Herzog, H. Zbinden, and N. Gisin, Phys. Rev. A \textbf{57}, 3229 (1998); A. Zeilinger
Rev. Mod. Phys. \textbf{71}, S288 (1999);
M.A. Rowe, D. Kielpinski, V. Meyer, C.A. Sackett, W.M. Itano, C. Monroe, and D.J. Wineland, Nature \textbf{409}, 791 (2001);
M. Giustina \emph{et al.}, Nature {\bf 497}, 227 (2013);
B. Hensen \emph{et al.}, Nature {\bf 526}, 682 (2015);
M. Giustina \emph{et al.}, Phys. Rev. Lett. {\bf 115}, 250401 (2015);
L. K. Shalm \emph{et al.}, Phys. Rev. Lett. {\bf 115}, 250402 (2015);
H.-P. Lo, C.-M. Li, A. Yabushita, Y.-N. Chen, C.-W. Luo, T. Kobayashi, Scientific Reports {\bf 6}, 22088 (2016).

\bibitem{CHSHineq} J.F. Clauser, M.A. Horne, A. Shimony, and R.A. Holt, 
Phys. Rev. Lett. \textbf{23}, 880 (1969).


\bibitem{Bell71} J.S. Bell, in \emph{Foundations of Quantum Mechanics}, ed. B. d'Espagnat (Academic, New York, 1971).

\bibitem{POMD} P. Hyllus, O. G{\"u}hne, D. Bru{\ss}, and M. Lewenstein
Phys. Rev. A \textbf{72}, 012321 (2005).


\bibitem{horo_bell} R. Horodecki, P. Horodecki, and M. Horodecki, Phys. Lett. A {\bf 200}, 340 (1995).

\bibitem{bennett_eof0} C. H. Bennett, G. Brassard, S. Popescu, B. Schumacher, J. Smolin, and W. K. Wootters, Phys. Rev. Lett 78, 2031 (1996).

\bibitem{bennett_eof1} C. H. Bennett, H. J. Bernstein, S. Popescu, and B. Schumacher, Phys. Rev. A {\bf 53}, 2046 (1996).

\bibitem{vedral_ent} V. Vedral, M. B. Plenio, M. A. Rippin, and P. L. Knight, Phys. Rev. Lett. {\bf 78}, 2275 (1997).

\bibitem{horo_ent_limit} M. Horodecki, P. Horodecki, and R. Horodecki, Phys. Rev. Lett. {\bf 84}, 2014 (2000).


\bibitem{horo_review} R. Horodecki, P. Horodecki, M. Horodecki, and K. Horodecki, Rev. Mod. Phys. {\bf 81}, 865 (2009).

\bibitem{uhlmann_convex} A. Uhlmann, Open Sys. Inf. Dyn. {\bf 5}, 209 (1998).

\bibitem{hill_conc}S. Hill and W. K. Wootters, Phys. Rev. Lett. {\bf 78}, 5022 (1997).

\bibitem{wootters_conc}W. K. Wootters, Phys. Rev. Lett. {\bf 80}, 2245 (1998).

\bibitem{hayden_cost}P. M. Hayden, M. Horodecki, and B. M. Terhal, J. Phys. A: Math. Gen. {\bf 34(35)}, 6891 (2001).

\bibitem{rains} E. M. Rains, Phys. Rev. A, {\bf 60}, 173 (1999).


\bibitem{plenio_ent} M. B. Plenio and S. Virmani, Quant. Inf. Comput. {\bf 7}, 1 (2007).

\bibitem{vidal_bound} G. Vidal and J. I. Cirac, Phys. Rev. Lett. {\bf 86}, 5803 (2001);
G. Vidal and J. I. Cirac, Phys. Rev. A {\bf 65}, 012323 (2001).


\bibitem{GM1} A. Shimony, Ann. N.Y. Acad. Sci. {\bf 755}, 675 (1995).

\bibitem{vedral_rel_ent} V. Vedral and M. B. Plenio, Phys. Rev. A {\bf 57}, 1619 (1998).



\bibitem{vedral_rmp_re} V. Vedral, Rev. Mod. Phys. {\bf 74}, 197 (2002).


\bibitem{NChuang}  M. Nielsen and I. Chuang, \textit{Quantum Computation and Quantum Information} (Cambridge University Press,
Cambridge, 2000), ISBN 9781139495486.



\bibitem{rel_ent_closed} A. Miranowicz and S. Ishizaka, Phys. Rev. A {\bf 78}, 032310 (2008);
						S. Friedland and G. Gour, J. Math. Phys. {\bf 52}, 052201 (2011).


\bibitem{karol_neg}  K. \.Zyczkowski, P. Horodecki, A. Sanpera, and M. Lewenstein, Phys. Rev. A {\bf 58}, 883 (1998).

\bibitem{eisert_neg}J. Eisert and M. B. Plenio, J. Mod. Opt. {\bf 46}, 145 (1999);
J. Lee, M. S. Kim, Y. J. Park, and S. Lee, J. Mod. Opt. {\bf 47}, 2151 (2000).

\bibitem{vidal_neg} G. Vidal and R.F. Werner, Phys. Rev. A {\bf 65}, 032314 (2002).

\bibitem{plenio_neg} M.B. Plenio, Phys. Rev. Lett. {\bf 95}, 090503 (2005).

\bibitem{Bennett-ek} C.H. Bennett, P.W. Shor, J.A. Smolin, and A.V. Thapliyal, Phys. Rev. Lett. \textbf{83}, 3081 (1999).

\bibitem{Bennett-dui} C.H. Bennett, P.W. Shor, J.A. Smolin, and A.V. Thapliyal, 
\emph{Entanglement-assisted capacity of a quantum channel and the reverse Shannon theorem},
quant-ph/0106052.

\bibitem{MarieCurie} M. Horodecki, P. Horodecki, R. Horodecki, D. Leung, and B. Terhal,
Q. Inf. and Comput. \textbf{1}, 70 (2001).

\bibitem{Debu} A. Winter,
J. Math. Phys. \textbf{43}, 4341 (2002).





\bibitem{ek1} S. Bose, M.B. Plenio, and V. Vedral, J. Mod. Opt. \textbf{47}, 291 (2000).

\bibitem{dui1} T. Hiroshima, J. Phys. A: Math. Gen. \textbf{34}, 6907 (2001). 

\bibitem{char1} X.S. Liu, G.L. Long, D.M. Tong, and F. Li, Phys. Rev. A \textbf{65}, 022304 (2002).


\bibitem{tin1} M. Ziman and V. Bu{\v z}ek, Phys. Rev. A \textbf{67}, 042321 (2003).



\bibitem {ref-halum}
J.P. Gordon, in \emph{Proc. Int. School Phys. ``Enrico Fermi,
Course XXXI''}, ed. P.A. Miles, p. 156 (Academic Press, NY 1964); 
L.B. Levitin, in \emph{Proc. VI National Conf. Inf. Theory, Tashkent}, p. 111
(1969); A.S. Holevo, Probl. Pereda. Inf. \textbf{9}, 3 1973 [Probl. Inf.
Transm. \textbf{9}, 110 (1973)].





\bibitem{Chennai}  T. M. Cover and J. A. Thomas, \emph{Elements of Information Theory} (Wiley, New York, 1991).

\bibitem{Utpakhi} R. Josza, D. Robb, and W.K. Wotters, Phys. Rev. A, \textbf{49}, 668 (1994).

\bibitem{Rajabazar1} B. Schumacher, M. Westmoreland, and W. K. Wootters,
Phys. Rev. Lett. \textbf{76}, 3452 (1996).

\bibitem{Khajuraho} P. Badzi{\c a}g, M. Horodecki, A. Sen(De),  and U. Sen, Phys. Rev. Lett. 
\textbf{91}, 117901 (2003).


\bibitem{Rajabazar2} M Horodecki, J. Oppenheim, A. Sen(De), and U. Sen,  Phys. Rev. Lett. \textbf{93}, 170503 (2004).




\bibitem{babarey} B. Schumacher and M.D. Westmoreland, Phys. Rev. A \textbf{56}, 131 (1997). 

\bibitem{maarey} A.S. Holevo, IEEE Trans. Inf. Theory \textbf{44}, 269 (1998).


\bibitem{TajMahal}  D. Bru{\ss}, G.M. D'Ariano, M. Lewenstein, C. Macchiavello, A. Sen(De), and U. Sen,
Phys. Rev. Lett. \textbf{93}, 210501 (2004).

\bibitem{ar-keu-korechhey?} D. Bru{\ss}, G.M. D'Ariano, M. Lewenstein, C. Macchiavello, A. Sen(De), and U. Sen, Int. J. Quant. Inf. {\bf 4}, 415 (2006). 



\bibitem{GHZ} D.M. Greenberger, M.A. Horne, and A. Zeilinger, in \emph{Bell's
 Theorem, Quantum Theory, and Conceptions of the Universe}, ed. M. Kafatos
 (Kluwer Academic,
 Dordrecht, 1989).


\bibitem{dur_vidal_cirac} W. D\"ur, G. Vidal, and J. I. Cirac, Phys. Rev. A {\bf 62}, 062314 (2000).


\bibitem{W_state}  A. Zeilinger, M. Horne, and D. Greenberger, in \emph{Squeezed
States and Quantum Uncertainty}, edited by D. Han, Y. S.
Kim, and W. W. Zachary (NASA Conference Publication
3135, NASA, College Park, 1992).



\bibitem{qc_wt_cc} D. Kaszlikowski, A. Sen(De), U. Sen, V. Vedral, and A. Winter, Phys. Rev. Lett. {\bf 101}, 070502 (2008).



\bibitem{GM2}  M. B. Plenio and V. Vedral,  J. Phys. A: Math. Gen. {\bf 34} 6997 (2001).

\bibitem{GM3} H. Barnum and N. Linden, J. Phys. A {\bf 34}, 6787 (2001);
								T.-C. Wei and P.M. Goldbart, Phys. Rev. A {\bf 68}, 042307 (2003).

\bibitem{GGM1} M. Blasone, F. Dell’Anno, S. De Siena, and F. Illuminati, Phys. Rev. A {\bf 77}, 062304 (2008).

\bibitem{GGM2} A. Sen(De) and U. Sen, Phys. Rev. A {\bf 81}, 012308 (2010); A. Sen(De) and U. Sen, \emph{Bound Genuine Multisite Entanglement: Detector of Gapless-Gapped Quantum Transitions in Frustrated Systems}, arXiv:1002.1253 [quant-ph].

\bibitem{GGM3} M. Cianciaruso, T. R. Bromley, and G. Adesso, Nature npj Quantum Information {\bf 2}, 16030 (2016).

\bibitem{GGM4} L. E. Buchholz, T. Moroder, and O. G\"uhne, Ann. Phys. (Berlin) {\bf 528}, 278 (2016).

\bibitem{GGM5} T. Das, S. S. Roy, S. Bagchi, A. Misra, A. Sen(De), U. Sen, Phys. Rev. A {\bf 94}, 022336 (2016).

\bibitem{slocc} C. H. Bennett, S. Popescu, D. Rohrlich, J. A. Smolin, and A. V. Thapliyal, Phys. Rev. A {\bf 63}, 012307 (2000).

\bibitem{frank_9} F. Verstraete, J. Dehaene, B. De Moor, and H. Verschelde, Phys. Rev. A {\bf 65}, 052112 (2002); D. Li, X. Li, H. Huang,
and X. Li, Phys. Lett. A {\bf 359}, 428 (2006);
L. Lamata, J. Le\'on, D. Salgado, and E. Solano, Phys. Rev. A {\bf 75}, 022318 (2007).


\bibitem{miyake} A. Miyake, 	Phys. Rev. A {\bf 67}, 012108 (2003).


\bibitem{ckw} V. Coffman, J. Kundu, and W. K. Wootters, Phys. Rev. A {\bf 61}, 052306 (2000).





\bibitem{terhal_mono} B. M. Terhal, \emph{Is Entanglement Monogamous?}, arxiv:quant-ph/0307120 (2013), and references therein.

\bibitem{mono_review} J. S. Kim, G. Gour, and B. C. Sanders,  Contemp. Phys. {\bf 53}, 417 (2012);
H. S. Dhar, A. K. Pal, D. Rakshit, A. Sen(De), and U. Sen,  arXiv:1610.01069 (2016).




\bibitem{frank_mono} T. J. Osborne and F. Verstraete, Phys. Rev. Lett. {\bf 96}, 220503 (2006).



\bibitem{mono_score} M. N. Bera, R. Prabhu, A. Sen(De), and U. Sen, Phys. Rev. A {\bf 86}, 012319 (2012).

















\bibitem{panch} W. D{\"u}r, J.I. Cirac, and R. Tarrach,
Phys. Rev. Lett. \textbf{83}, 3562 (1999). 

\bibitem{saat} R.F. Werner and M.M. Wolf,
Phys. Rev. A \textbf{64}, 032112 (2001); M. \.Zukowski and {\v C}. Brukner,
Phys. Rev. Lett. \textbf{88}, 210401 (2002)


\bibitem{Mumbai}  G. Svetlichny, Phys. Rev. D \textbf{35}, 3066 (1987); D. Collins, N. Gisin, S. Popescu, D. Roberts, and V. Scarani
Phys. Rev. Lett. \textbf{88}, 170405 (2002); M. Seevinck and 
G. Svetlichny,  Phys. Rev. Lett. \textbf{89}, 060401 (2002); S.M. Roy, Phys. Rev. Lett. \textbf{94}, 010402 (2005). 


\bibitem{satattor} A. Ac{\'i}n, D. Bru{\ss}, M. Lewenstein, and A. Sanpera,
Phys. Rev. Lett. \textbf{87}, 040401 (2001).

\bibitem{Horo-realign} M. Horodecki, P. Horodecki, and R. Horodecki, Open Sys. Inf. Dyn. {\bf 13}, 103 (2006).


\bibitem{ek} C. Moura Alves and D. Jaksch
Phys. Rev. Lett. \textbf{93}, 110501 (2004).
 
\bibitem{chhoi} W. Laskowski, T. Paterek, M. \.Zukowski, and {\v C}. Brukner,
Phys. Rev. Lett. \textbf{93}, 200401 (2004).



\bibitem{char} F.G.S.L. Brandao and R.O. Vianna,
Phys. Rev. Lett. \textbf{93}, 220503 (2004). 
 

\bibitem{tin} G. Toth and O G{\"u}hne,
Phys. Rev. Lett. \textbf{94}, 060501 (2005).

\bibitem{dui} A.C. Doherty, P.A. Parrilo, and F.M. Spedalieri,
Phys. Rev. A \textbf{71}, 032333 (2005).

\bibitem{sotero} O. G{\"u}hne, G. Toth, and H.J. Briegel, New J. Phys. {\bf 7}, 229 (2005). 













\bibitem{UPBPRL} C.H. Bennett, D.P. DiVincenzo, T. Mor, P.W. Shor, J.A. Smolin, and B.M. Terhal, Phys. Rev. Lett. \textbf{82}, 5385 (1999).


\end{thebibliography}
\end{document}